\documentclass[pdftex,twocolumn,epjc3]{svjour3}          

\RequirePackage[T1]{fontenc}

\smartqed  

\RequirePackage{graphicx}
\RequirePackage{mathptmx}      
\RequirePackage{flushend}
\RequirePackage[numbers,sort&compress]{natbib}
\RequirePackage[colorlinks,citecolor=blue,urlcolor=blue,linkcolor=blue]{hyperref}

\journalname{Eur. Phys. J. C}

\usepackage{amsmath}
\usepackage{amssymb}
\usepackage{tensor}

\newcommand*{\bfrac}[2]{\genfrac{}{}{0pt}{}{#1}{#2}}
\newcommand*{\sgn}{\mathrm{sgn}}

\begin{document}

\title{Constructing spherically symmetric Einstein-Dirac systems with multiple spinors: Ansatz, wormholes and other analytical solutions}


\author{Jose Luis Bl\'azquez-Salcedo\thanksref{e1,addr1}
	         \and
        Christian Knoll\thanksref{e2,addr1}
}

\thankstext{e1}{e-mail: jose.blazquez.salcedo@uni-oldenburg.de}  

\thankstext{e2}{e-mail: c.knoll110@gmail.com} 


\institute{Institut f\"ur Physik, Universit\"at Oldenburg, D-26111 Oldenburg, Germany\label{addr1}}

\date{Received: date / Accepted: date}

\maketitle

\begin{abstract}
In this paper we present a detailed calculation of an Ansatz that allows to obtain spherically symmetric Einstein-Dirac configurations in $d$-dimensions. We show that this is possible by combining $2^{\lfloor \frac{d-2}{2} \rfloor}$ Dirac fields, making use of the properties of the angular dependence of the spinors in a spherical background.
By applying this Ansatz, we investigate some simple analytical solutions. One of them is a regular wormhole supported by the Dirac fields. Other solutions include a pathological black hole and a naked singularity. We analyze the domain of existence and properties of all these solutions.
\end{abstract}

\section{Introduction}

The study of the Einstein equations coupled to different classes of matter content has received a lot of attention in the recent years, since solutions of these theoretical models could be related to exotic astrophysical systems (composing the dark matter/energy sector) \cite{Barack:2018yly}. In higher dimensions these solutions could be of potential interest in the context of supergravity and the AdS/CFT correspondence \cite{Ammon:2015wua}.  

There is an ongoing intense exploration of self-\-gra\-vitating stationary soliton-like solutions composed by different classes of massive fundamental fields. 
The interest of these settings is because of the contrast with the more standard Einstein-Maxwell theory (and even with the Einstein-Maxwell-scalar theory \cite{Herdeiro:2019oqp}), where the electro-vac black hole is the only self-gravitating stationary soliton-like solution. But when fundamental fields are considered to be massive the situation changes. It is possible to construct particle like solutions (with regular, stationary space-times, but typically with a harmonic time dependence on the fields).

With scalar fields (in particular, massive and complex), these configurations are well-known to exist and originally obtained in \cite{PhysRev.172.1331,PhysRev.187.1767}. Typically these configurations are known as boson stars \cite{Schunck:2003kk}, and they are considered potential candidates as astrophysical objects \cite{Liebling:2012fv}.
With massive vector fields (known as Proca stars) this was explored in \cite{Brito:2015pxa} (some results in five dimensions can be found in \cite{Duarte:2016lig}). Several astrophysical properties of these objects have been analyzed in depth \cite{Cunha:2017wao,Sanchis-Gual:2017bhw,Shen:2016acv,Cao:2016zbh}.

From a more theoretical point of view, a particularly interesting case is to consider self-gravitating soliton-like solutions of the Einstein-Dirac system, where gravity plays the role of the non-linear interaction that allows for the existence of Dirac solitons in simpler models
\cite{Soler:1970xp,Ranada:1984fv}. However, there is an additional challenge for this type of fields. Because of the intrinsic angular momentum of a single spinor field (a preferred direction in space-time), the resulting space-time is forced to rotate in order to accommodate stationary solutions. Such configurations have been very recently constructed in \cite{Herdeiro:2019mbz}, where they were compared with rotating Boson and Proca stars.

Nonetheless, a possible route to enforce that the global solution of the Einstein-Dirac system is actually static and spherically symmetric, is to relax the single spinor condition and consider a collection of Dirac fields. 
In four dimensions, two fields are enough to cancel the intrinsic angular momentum and realize a global spherically symmetric and non-rotating space-time \cite{Finster:1998ws,Herdeiro:2017fhv} (provided the fields possess a certain harmonic time dependence).

The properties of such multi-Dirac soliton-like solutions have been recently studied in \cite{Blazquez-Salcedo:2019qrz}, where their properties were compared with similar configurations made of scalar and Proca fields for $d \ge 4$. It was shown that some generic features of the solutions actually do not depend qualitatively on the spin of the field, but they are controlled by  the dimension of the space-time. These 'Dirac stars' have also been studied in the presence of vector fields \cite{Dzhunushaliev:2019kiy}.

On the other hand, solutions of the Einstein-Dirac system with multiple fermions are of interest in condensed matter \cite{Gonzalez:1992qn}, where in particular wormhole space-times with two Dirac fields can be used as effective models describing two graphene layers connected by a short nanotube. This model is known as the graphene wormhole \cite{Gonzalez:2009je,Atanasov_2011,Pincak2013,Smotlacha:2014tza,Sepehri:2016svv,Sepehri:2017dky,Garcia:2019gro, Cariglia:2018rhw, Rojjanason:2018icy,Maldacena:2018gjk}

The purpose of the present paper is two-fold. First, we want to provide details on how the Ansatz for these multi-Dirac self-gravitating solitons is calculated for arbitrary space-time dimension. Second, we will show that, in addition to the numerical solutions previously obtained in \cite{Blazquez-Salcedo:2019qrz}, it is possible to obtain a few simple analytical solutions to the equations. One of these solutions is a regular wormhole solution supported by multiple Dirac fields. We will analyze the physical meaning and properties of these configurations.

The paper is organized as follows: in section \ref{sec:vielbein} we present the general formalism for the Einstein-Dirac system and make an overview of how the Ansatz is built. In section \ref{section_full_ansatz} we explain how to combine the angular dependence of the different Dirac fields in order to get a spherically symmetric stress-energy tensor. In section \ref{section_action} we analyze the effective action and the minimum set of differential equations of this system. In section \ref{section_sols} we describe several sets of solutions that can be obtained in various particular cases.
In section \ref{section_wormhole} we present a regular wormhole solution supported by pairs of Dirac fields.
 In section \ref{bh_section} we present a black hole solution with a pathological behaviour of the Dirac fields at the horizon.  In section \ref{sing_section} we present a light-like singularity. In section \ref{section_conclusions} we end the paper with a summary and conclusions.

\section{Overview of the general setting}\label{sec:vielbein}

We want to construct spherically symmetric solutions of the $d$-dimensional Einstein-Dirac system.
There are many studies on the Dirac equation in higher dimensional, spherically symmetric space-times \cite{Dong:2011zzf,Dong:2003xy,Dong:2003wi,Cotaescu:2003be,Chakrabarti:2008xz,Sporea:2015wsa,Blazquez-Salcedo:2017bld,Gonzalez:2018xrq,Sporea:2019iwk,Bronnikov:2019nqa}.
 An appropriate metric Ansatz is
\begin{eqnarray} \label{eqn:spherically_spacetime}
\mathrm d s^2 &=& N(r) \sigma^2(r) \, \mathrm d t^2 - \frac{1}{N(r)} \, \mathrm d r^2 - r^2 \, \mathrm d \Omega^2_{d-2} \, ,
\end{eqnarray}
where $\mathrm d \Omega_{d-2}^2$ is the line element of the $(d-2)$-sphere. 
In order to build minimally coupled Dirac fields to this metric, we need to specify the vielbein
, and for the metric Ansatz (\ref{eqn:spherically_spacetime}) we choose
\begin{eqnarray}\label{eqn:Vielbein}
\boldsymbol{\omega}^t &=& \sqrt{N} \sigma \mathbf d t \, , \nonumber \\
\boldsymbol{\omega}^r &=& \frac{1}{\sqrt{N}} \mathbf d r \, ,  \\
\boldsymbol{\omega}^j &=& r \, \boldsymbol{\omega}^j_{d-2} \, , \nonumber
\end{eqnarray}
where $j = 1, ..., d-2$ is an index running over the $(d-2)$-sphere and $\boldsymbol{\omega}^j_{d-2}$ is a vielbein for the $(d-2)$-sphere. This allows us to write the Dirac equation,
\begin{eqnarray}
\label{eqn:Dirac}
\mathcal D \Psi &=& \left[ \frac{\mathrm i}{\sqrt{N} \sigma} \gamma^t \partial_t + \mathrm i \sqrt{N} \gamma^r \left( \partial_r + \frac{\mathrm d}{\mathrm d r} \ln \sqrt{ \sqrt{N} \, \sigma r^{d-2}} \right) \right. \nonumber \\
&&\left. + \frac{\mathrm i}{r} \gamma^t \gamma^r \mathcal K_{d-2} - m \right] \Psi = 0 \, .
\end{eqnarray}
The operator $\mathcal K_{d-2}$ is the angular operator of the $(d-2)$-sphere, given by
\begin{eqnarray}
\mathcal K_{d-2} = \gamma^j_{d-2} \mathbf e_j^{d-2} + \frac{1}{2} \Gamma^{d-2}_{|ij|}(\mathbf e_k^{d-2}) \gamma_{d-2}^k \gamma_{d-2}^i \gamma_{d-2}^j \, , 
\end{eqnarray}
with $\Gamma^{d-2}_{ij}$ being the spin connection of the $(d-2)$-sphere, $\mathbf e_j^{d-2}$ being the dual to the vielbein on the $(d-2)$-sphere and $\gamma^j_{d-2} = \gamma^t \gamma^r \gamma^j$. Because of the spherical symmetry of the metric, the Dirac operator commutes with the angular operator, $[\mathcal D, \mathcal K_{d-2}]=0$. In addition $\partial_t$ is a Killing vector. These two properties allow us to write a spinor with the following Ansatz
\begin{eqnarray}
\label{eqn:spinor1}
\Psi = \mathrm e^{-\mathrm i \omega t} \phi_\kappa(r) \otimes \Theta_\kappa \, ,
\end{eqnarray}
where $\Theta_\kappa$ depends on the angular variables only. The angular part is chosen to fulfill $\mathcal K_{d-2} \Theta_\kappa = \kappa \Theta_\kappa$, with $\kappa$ the angular momentum eigenvalue. 

As we said in the introduction, because of a single spinor having a non-trivial intrinsic angular momentum, it is not possible to construct a compatible solution of the Dirac equation (\ref{eqn:Dirac}) with the metric (\ref{eqn:spherically_spacetime}). 

A way out is to consider a system of multiple Dirac spinors. If we choose them appropriately, the combination of all of them will have a total stress-energy tensor compatible with the symmetries of the metric (\ref{eqn:spherically_spacetime}).
To do so, we need $2^{\lfloor \frac{d-2}{2} \rfloor}$ spinors (note $\lfloor x \rfloor$ means the integer less than or equal to $x$).  These spinors need to have the same radial function, but they differ in their angular parts, with the same (the smallest possible) angular eigenvalue combined incoherently. This means that, written as a formal sum, the spinors combine like  
\begin{eqnarray}
\mathrm e^{-\mathrm i \omega t} \bigoplus_{\boldsymbol \epsilon} \phi_\kappa \otimes \Theta_{\kappa, \boldsymbol \epsilon} \, , \\ \nonumber
\end{eqnarray}
where $\epsilon$ is the index of each one of the $2^{\lfloor \frac{d-2}{2} \rfloor}$ spinors. 

In terms of the action, the Einstein-Dirac system with cosmological constant $\Lambda$ can be written like
\begin{eqnarray}
\label{eqn:action_Diracs}
S &=&  \int \mathrm d^d x \sqrt{|g|} \, \left[ R + \frac{2\alpha_g}{\sqrt{|g|}}\mathcal L_\text{spinors} + \Lambda \ \right] \, ,
\end{eqnarray}
where $\alpha_g$ is the coupling constant between gravity and the spinor fields, and the Lagrangian for the spinor part is then a sum of the form
\begin{eqnarray}
\label{eqn:Lagrangian_Diracs}
\mathcal L_\text{spinors} = 
\sum\limits_{\boldsymbol{\epsilon}} \left[ \frac{\mathrm i}{2} \overline{\Psi}_{\boldsymbol{\epsilon}} \gamma^a \nabla_a \Psi_{\boldsymbol \epsilon} - \frac{\mathrm i}{2} \nabla_a \overline{\Psi}_{\boldsymbol{\epsilon}}  \gamma^a \Psi_{\boldsymbol{\epsilon}}  - m \overline{\Psi}_{\boldsymbol{\epsilon}}  \Psi_{\boldsymbol{\epsilon}}  \right] \, . 
\end{eqnarray}
We will show that this leads to a spherically symmetric energy momentum tensor for the spinors,
\begin{eqnarray}
\label{eqn:s-e-tensor}
T_{\mu \nu} = \sum\limits_{\boldsymbol \epsilon} T^{(\boldsymbol \epsilon)}_{\mu \nu} = 2 \sum\limits_{\boldsymbol{\epsilon}} \Im ( \overline{\Psi}_{\boldsymbol{\epsilon}} \gamma_{(\mu} \nabla_{\nu)} \Psi_{\boldsymbol{\epsilon}} ) \, ,
\end{eqnarray}
where the $T^{(\boldsymbol \epsilon)}_{\mu \nu}$ is the energy momentum tensor of the $\mathrm e^{-\mathrm i \omega t}  \phi_\kappa \otimes \Theta_{\kappa, \boldsymbol \epsilon}$ spinor. 

We will now focus on the $(d-2)$-dimensional sphere and the construction of this spherically symmetric configuration.

\section{How to combine the spinors}
\label{section_full_ansatz}

In the following we will make use of expression (\ref{eqn:spinor1}) for each one of the spinors. We will assume that: 
\begin{enumerate}
\item  all the spinors share the same radial dependence;
\item all the spinors share the same temporal dependence, and we will assume it can be written in terms of a phase, introducing the frequency $\omega$; 
\item the spinors only differ in the angular part.
\end{enumerate}

In this section we will discuss in detail the properties of this angular part, and how it can be chosen in order to make the stress-energy tensor compatible with spherical symmetry.

\subsection{Peeling the $n$-sphere}

Let $n$ denote the dimension of the sphere. Solutions to the Dirac equation on the $n$-sphere are well-known and called spinor monopole harmonics in the literature \cite{Wu:1976ge,Pereira:1982kv,Camporesi:1995fb}. It is however not trivial to combine these solutions into a field configuration which possesses a spherically symmetric energy momentum tensor. 
An approach to make this combination more intuitive, is to 
maximize the commuting Killing vectors on the $n$-sphere in our coordinate system. For this we will choose angular coordinates in such a way, that the line element on the sphere is given recursively by
\begin{eqnarray}
\mathrm d \Omega_n^2 = \mathrm d \theta_n^2  + S_n^2 \, \mathrm d \phi_n^2 + C_n^2 \, \mathrm d \Omega_{n-2}^2 \, , \nonumber \\
\end{eqnarray}
with
\begin{eqnarray}
\mathrm d \Omega_{n-2}^2 = \begin{cases} 0 & \text{, } n = 2 \\ \mathrm d \phi_1^2 & \text{, } n = 3 \\ (n-2) \ \text{sphere line element}&\text{, } n > 3 \, \end{cases} 
\end{eqnarray}
and $S_n = \sin \theta_n$, $C_n = \cos \theta_n$. We thus slice off a two-sphere from the $n$-sphere. This is convenient, because it allows us to define the vielbein on the sphere also in a recursive way, meaning
\begin{eqnarray}
\boldsymbol{\omega}^{\theta_n}_n = \mathbf d \theta_n \, , \nonumber \\
\boldsymbol{\omega}^{\phi_n}_n = S_n \, \mathbf d \phi_n \, , \\
\boldsymbol{\omega}^j_n = C_n \, \boldsymbol{\omega}^j_{n-2} \, , \nonumber
\end{eqnarray}
with $j = 1, \dots, n-2$ being an index running over the $(n-2)$-sphere. The spinor covariant derivative on the $n$-sphere $\nabla_a^{(n)}$ is thus
\begin{eqnarray}
\nabla^{(n)}_{\theta_n} &=& \partial_{\theta_n} \, , \nonumber \\
\nabla^{(n)}_{\phi_n} &=& \frac{1}{S_n} \partial_{\phi_n} - \frac{C_n}{2 S_n} \gamma_n^{\mathcal K_n} \, ,  \\
\nabla^{(n)}_j &=& \frac{1}{C_n} \nabla_j^{(n-2)} - \frac{\mathrm i S_n}{2 C_n} \gamma^{\phi_n}_n \gamma_{n-2}^j \, , \nonumber
\end{eqnarray}
with $\gamma^{\mathcal K_n}_n = - \gamma^{\theta_n}_n \gamma^{\phi_n}_n$ and $\gamma_{n-2}^j = \mathrm i \gamma^{\mathcal K_n}_n \gamma^j_n$ and $j$ as before. One can think of $\gamma_{n-2}^j$ as the $\gamma_n^j$ matrices projected down onto the $(n-2)$-sphere with $\gamma^{\mathcal K_n}_n$ governing this projection. The reason we choose this factorization of the $\gamma$-matrices will become clear later. 

With these choices of line element, vielbein and algebra, we can write the Dirac operator on the sphere (meaning the angular part) as 
\begin{eqnarray}
\mathcal K_n &=&  \gamma_n^a \nabla^{(n)}_a \nonumber \\
&=& \gamma^{\theta_n}_n \left( \partial_{\theta_n} + \partial_{\theta_n} \ln \sqrt{S_n \, C^{n-2}_n} \right)  \\
&&+ \frac{1}{S_n} \gamma_n^{\phi_n} \partial_{\phi_n} + \frac{\mathrm i}{C_n} \gamma^{\mathcal K_n}_n \mathcal K_{n-2} \, , \nonumber
\end{eqnarray}
where the index $a$ runs over the $n$-sphere and $\mathcal K_{n-2} = \gamma_{n-2}^k \nabla_k^{(n-2)}$ is the angular operator for the $(n-2)$-sphere.

The matrices fulfill
\begin{eqnarray}
\left\{ \gamma^a_n, \gamma^b_n \right\} &=& - 2 \delta^{a b} \, , \text{ with } a, b \in \{\theta_n, \phi_n, j \} \, , \label{eqn:Clifford-n-sphere} \\
\left\{ \gamma^a_n, \gamma^b_n \right\} &=& - 2 \delta^{a b} \, , \text{ with } a, b \in \{\theta_n, \phi_n, \mathcal K_n \} \, , \;\; \label{eqn:Clifford-2-sphere}  \\
\left\{ \gamma_{n-2}^j, \gamma_{n-2}^k \right\} &=& -2 \delta^{j k} \, , \label{eqn:Clifford-n-2-sphere} \\
\left[ \gamma^{\mathcal K_n}_n, \gamma^j_n \right]  &=& 0   \, , \label{eqn:projector-commute} \\
\left[ \gamma^a_n, \gamma^j_{n-2} \right] &=& 0 \, , \text{ with } a \in \{\theta_n, \phi_n, \mathcal K_n\} \, , \label{eqn:angular-projected} 
\end{eqnarray}
where in the above $j$ and $k$ denote indices on the $(n-2)$-sphere. 

Equation (\ref{eqn:Clifford-n-sphere}) expresses the Clifford algebra on the $n$-sphere. Equation (\ref{eqn:Clifford-2-sphere}) is the Clifford algebra on the 2-sphere. Finally equation (\ref{eqn:Clifford-n-2-sphere}) is the Clifford algebra on the $(n-2)$-sphere, showing us that the projection works correctly and we have sliced off a two sphere from the $n$-sphere. The next equation (\ref{eqn:projector-commute}) tells us that the matrix governing the projection onto the $(n-2)$-sphere commutes with the matrices of the $(n-2)$-sphere. The last equation (\ref{eqn:angular-projected}) shows that the projected $\gamma$-matrices on the $(n-2)$-sphere commute with the $\gamma$-matrices on the sliced off two-sphere. 

This last relation implies that $[\mathcal K_n, \mathcal K_{n-2}] = 0$.  In addition, since $\partial_{\phi_n}$ is a Killing vector we also have $[\partial_{\phi_n}, \mathcal K_n] = 0$. Even more, for any $n, m \in \mathbb N$ we have that in the tower of angular operators $[\mathcal K_n, \mathcal K_m] = 0$ and $[\mathcal K_n, \partial_{\phi_m}] = 0$.

\subsection{Angular solutions of the spinor field}

We have rewritten the angular operator into a tower of angular operators. We will study now what this is implying to the angular part of the spinor fields.

Denote by $\Theta_{\kappa_n}$ the eigenspinor $\mathcal K_n \Theta_{\kappa_n} = \kappa_n \Theta_{\kappa_n}$. Due to the above comutator (\ref{eqn:angular-projected}), we can factorize the angular part of the solution and write
\begin{eqnarray}
\Theta_{\kappa_n} = \mathrm e^{\mathrm i m_n \phi_n} \Theta_{\kappa_n, m_n} \otimes \Theta_{\kappa_{n-2}} \, ,
\end{eqnarray}
with $\mathcal K_{n-2} \Theta_{\kappa_{n-2}} = \kappa_{n-2} \Theta_{n-2}$. This leads to the equation for $\Theta_{\kappa_n, m_n}$
\begin{eqnarray}\label{eqn:Diff_Theta_kappa_n}
\left[ \gamma^{\theta_n}_n \left( \frac{\mathrm d}{\mathrm d \theta_n} +\frac{\mathrm d}{\mathrm d \theta_n} \sqrt{S_n \, C^{n-2}_n} \right)  + \frac{\mathrm i m_n}{S_n} \gamma_n^{\phi_n}  + \frac{\mathrm i \kappa_{n-2}}{C_n} \gamma^{\mathcal K_n}_n \right] \Theta_{\kappa_n, m_n} \nonumber \\
= \kappa_n \Theta_{\kappa_n, m_n} \, . \;\;\;\;\;\;\;\;\;\; 
\end{eqnarray}
Let us study this equation a bit more. It is convenient to define
\begin{eqnarray}
\Theta_{\kappa_n, m_n} = \frac{\mathrm e^{- \frac{\theta_n}{2} \gamma_n^{\phi_n} \gamma_n^{\mathcal K_n}}}{\sqrt{S_n \, C^{n-2}_n}} \hat \Theta_n \, .
\end{eqnarray}
Substituting this into the differential equation (\ref{eqn:Diff_Theta_kappa_n}) and multiplying with $\mathrm e^{\frac{\theta_n}{2} \gamma_n^{\phi_n} \gamma_n^{\mathcal K_n}}$ from the left gives the following differential equation for $\hat \Theta_n$
\begin{eqnarray}
&&\left[ \gamma_n^{\theta_n} \frac{\mathrm d}{\mathrm d \theta_n} + \mathrm i \gamma^{\phi_n}_n \left( m_n \, \frac{C_n}{S_n} - \kappa_{n-2} \frac{S_n}{C_n} \right) \right. \nonumber \\
&&\left. + \mathrm i \gamma_n^{\mathcal K_n} \left(  m_n + \kappa_{n-2} \right) - \frac{1}{2} \gamma^{\theta_n}_n \gamma^{\phi_n}_n \gamma^{\mathcal K_n}_n \right] \hat \Theta_n = \kappa_n \hat \Theta_n \, . 
\end{eqnarray}

At this stage, let us choose as a representation
\begin{eqnarray}
\gamma_n^{\theta_n} &=& \left[ \begin{array}{cc} 0 & 1 \\ -1 & 0 \end{array} \right] \; , \; \gamma^{\phi_n}_n = \left[ \begin{array}{cc} 0 & \mathrm i \\ \mathrm i & 0 \end{array} \right] \, , \nonumber \\
\gamma^{\mathcal K_n}_n &=& \left[ \begin{array}{cc} - \mathrm i & 0 \\ 0 & \mathrm i \end{array} \right] \; , \; \hat \Theta_n = \left[ \begin{array}{c} \Theta_1 \\ \Theta_2 \end{array} \right] \, .
\end{eqnarray}
This leads to the coupled first order differential equation
\begin{eqnarray}\label{eqn:angular_diff}
\left( \frac{\mathrm d}{\mathrm d \theta_n} + m_n \frac{C_n}{S_n} - \kappa_{n-2} \frac{S_n}{C_n} \right) \Theta_1 &=& - K_+ \Theta_2 \, , \nonumber \\
\left( \frac{\mathrm d}{\mathrm d \theta_n} - m_n \frac{C_n}{S_n} + \kappa_{n-2} \frac{S_n}{C_n} \right) \Theta_2 &=& + K_- \Theta_1 \, , 
\end{eqnarray}
with
\begin{eqnarray}
K_\pm = \frac{1}{2} + \kappa_n \pm \left( m_n + \kappa_{n-2} \right) \, .
\end{eqnarray}
Let us also define
\begin{eqnarray}
p_m &:=& \left| m_n + \frac{1}{2} \right| \; , \; p_\kappa := \left| \kappa_{n-2} + \frac{1}{2} \right| \, , \nonumber \\
\mathcal F_j &:=& F \left( \bfrac{j + 1 - n_\kappa, j + n_\kappa + p_m + p_\kappa}{j + 1 + p_\kappa} ; C^2 \right) \, ,  \\
\mathcal R_j &:=& \frac{(j + 1 - n_\kappa) \left(j + n_\kappa + p_m + p_\kappa \right)}{j + 1 + p_\kappa} \, , \nonumber
\end{eqnarray}
where $F(a, b; c; z)$ is the hypergeometric function and $n_\kappa \ge 1$ is a natural number. For solutions of the differential equation (\ref{eqn:angular_diff}) $m_n$ is a half integer number, while $|\kappa_{n-2}| \ge (n-2)/2$ is an integer number in the case $n$ is even, or a half integer number in the case $n$ is odd. 

With these definitions, we can write three solutions of equation (\ref{eqn:angular_diff}).

1) The solution in the case $K_+ \neq 0$ is
\begin{eqnarray}
\Theta_1 &=& C_n^{p_2 + 1/2} \, S_n^{p_1 + 1/2} \, \mathcal F_0 \, , \nonumber \\
\Theta_2 &=& \left\{ \frac{2 C_n S_n \, \mathcal R_0 \mathcal F_1}{\mathcal F_0} - \left( m_n + \frac{1}{2} + p_1 \right) \frac{C_n}{S_n} \right. \, ,  \\
&&\left. \;\; + \left( \kappa_{n-2} + \frac{1}{2} + p_2 \right) \frac{S_n}{C_n} \right\} \frac{\Theta_1}{K_+} \nonumber
\end{eqnarray}
and the angular eigenvalue in this case is
\begin{eqnarray}
\kappa_n &=& - \frac{1}{2} \pm_\kappa \left| 2 n_\kappa - 1 + p_1 + p_2 \right|
\end{eqnarray}
where $\pm_\kappa$ is a sign choice.

2) In the case of $K_+ = 0$ the solution is
\begin{eqnarray}
\Theta_1 &=& 0 \, , \nonumber \\
\Theta_2 &=& S_n^{m_n} C_n^{\kappa_{n-2}} \, ,
\end{eqnarray}
with $m_n \ge 1/2$ and $\kappa_{n-2} \ge (n-2)/2$. The angular eigenvalue in this case is
\begin{eqnarray}
\kappa_n = -\frac{1}{2} - m_n - \kappa_{n-2} \, .
\end{eqnarray}

3) Lastly in the case of $K_- = 0$ the solution is
\begin{eqnarray}
\Theta_1 &=& S_n^{-m_n} C_n^{-\kappa_{n-2}} \, , \nonumber \\
\Theta_2 &=& 0 \, ,
\end{eqnarray}
with $m_1 \le - 1/2$ and $\kappa_{n-2} \le -(n-2)/2$. The angular eigenvalue in this case is
\begin{eqnarray}
\kappa = -\frac{1}{2} + m_n + \kappa_{n-2} \, .
\end{eqnarray}

With these three solutions at hand, let us analyze what are the smallest possible eigenvalues $|\kappa_n|$ we can reach and what are the corresponding angular solutions. An inspection of the solutions allows us to conclude that these are given by the cases $K_\pm = 0$ choosing $|m_n| = 1/2$, $|\kappa_{n-2}| = (n-2)/2$,  and by $K_+ \neq 0$ choosing $n_\kappa = 1$, $m_n = \pm 1/2$, $\kappa_{n-2} = \mp (n-2)/2$. This gives as a result $|\kappa_n|=n/2$.

Note that the previous values of the angular parameters depend only on $n$ and some possible sign choices. In the following we will choose only these minimum values.
Hence, since for a given dimension $n$ the only possible choices are the different signs of the angular parameters, for the sake of simplicity we can relabel the angular solution accordingly: $\Theta_{\kappa_n, m_n} \equiv \Theta_{\sgn (\kappa_n), \sgn( m_n)}^{(n)}$.
Note that the sign of $\kappa_{n-2}$ is determined by these sign choices via 
\begin{eqnarray}
\sgn (\kappa_{n-2}) = - \sgn( m_n )\, \sgn (\kappa_n) \, . 
\end{eqnarray}

Let us write explicitly what these solutions are. There are four possibilities with $|\kappa_n|=n/2$:
\begin{eqnarray}
\Theta^{(n)}_{++} &=& \left[ \begin{array}{c} \sin \frac{\theta_n}{2} \\ -\cos \frac{\theta_n}{2} \end{array} \right] \, , \, \Theta^{(n)}_{+-} = \left[ \begin{array}{c} \cos \frac{\theta_n}{2} \\ \sin \frac{\theta_n}{2} \end{array} \right] \, , \nonumber \\
\Theta^{(n)}_{-+} &=& \left[ \begin{array}{c} \sin \frac{\theta_n}{2} \\ \cos \frac{\theta_n}{2} \end{array} \right] \, , \, \Theta^{(n)}_{--} = \left[ \begin{array}{c} \cos \frac{\theta_n}{2} \\ -\sin \frac{\theta_n}{2} \end{array} \right] \, .
\end{eqnarray}

Tracing back our steps we can thus write the angular part of the spinor for the $n$-sphere with minimal absolute value of the eigenvalue $|\kappa_n|=n/2$ as
\begin{eqnarray}
\label{eqn:Theta_min}
\Theta_{\epsilon_\kappa, \boldsymbol{\epsilon}} \equiv \Theta_{\kappa_n} = \left[ \prod\limits_{0 < j \le n}^{j \equiv n \, \mathrm{mod} \, 2} \mathrm e^{\mathrm i \epsilon_j \phi_j / 2} \right] \bigotimes\limits_{1 < j \le n}^{j \equiv n \, \mathrm{mod} \, 2} \Theta^{(j)}_{\pm_j, \epsilon_j} \, , 
\end{eqnarray}
with $\epsilon_\kappa$ being the sign choice for $\kappa_n$, and the $\epsilon_j$ being the sign choices for the $m_j$ ($j>1$), which we can summarize as a binary vector $\boldsymbol \epsilon$ (notice that either the even or the odd components of this vector are immaterial for us depending on $n$), and
\begin{eqnarray}
\pm_j &=& \sgn(\kappa_j) = - \sgn (m_{j+2}) \sgn(\kappa_{j+2}) \nonumber \\
&=& \left[ \prod\limits_{j < k \le n}^{k \equiv n \, \mathrm{mod} \, 2} -\epsilon_k \right] \epsilon_\kappa \, .
\end{eqnarray}
The sign of $m_1$, so $\epsilon_1$, is fixed by the equation
\begin{eqnarray}
\epsilon_1 = \pm_1 = \left[ \prod\limits_{1 < k \le n}^{k \equiv n \, \mathrm{mod} \, 2} -\epsilon_k \right] \epsilon_\kappa \, .
\end{eqnarray}
The reason for this is that $m_1$ plays a double role as an eigenvalue to $\partial_{\phi_1}$ and as the angular eigenvalue $\kappa_1$ to the one-sphere (circle).

\subsection{Analyzing the properties of the angular solution on the components of the total stress-energy tensor} \label{s-e-tensor_components}

Now that we have this set of solutions (\ref{eqn:Theta_min}) for the angular part of the spinor, we need to study how it enters into the stress-energy tensor. The  stress-energy tensor for a collection of spinors was given in expression (\ref{eqn:s-e-tensor}). From there we can see that it is useful to construct explicitly the matrix elements of the covariant derivative $\nabla^{(n)}_a$ multiplied with a matrix $\Gamma$, since we will need these objects for the calculation of the total stress-energy tensor.

The first thing to do is to look at the following relations
\begin{eqnarray}
\gamma^{\theta_n}_n \Theta^{(n)}_{\pm_\kappa, \pm_m} &=& - \pm_\kappa \pm_m \Theta^{(n)}_{\pm_\kappa, \mp_m} \, , \nonumber \\
\gamma^{\phi_n}_n \Theta^{(n)}_{\pm_\kappa, \pm_m} &=& -\pm_\kappa \pm_m \Theta^{(n)}_{\mp_\kappa, \mp_m} \, , \nonumber \\
\gamma^{\mathcal K_n}_n \Theta^{(n)}_{\pm_\kappa, \pm_m} &=& - \mathrm i \Theta^{(n)}_{\mp_\kappa, \pm_m} \, , \nonumber \\
\partial_{\theta_n} \Theta^{(n)}_{\pm_\kappa, \pm_m} &=& \pm_m \Theta^{(n)}_{\pm_\kappa, \mp_m} / 2 = \mp_\kappa \gamma^{\theta_n}_n \Theta^{(n)}_{\pm_\kappa, \pm_m} / 2 \, .  
\end{eqnarray}
Using this and the inner product table
\begin{eqnarray}
\begin{array}{c|cccc}
 & \Theta_{+, +}^{(n)} & \Theta_{+, -}^{(n)} & \Theta_{-, +}^{(n)} & \Theta_{-, -}^{(n)} \\ \hline
\Theta_{+, +}^{(n) \, \dagger} & 1 & 0 & - C_n & S_n \\
\Theta_{+, -}^{(n) \, \dagger} & 0 & 1 & S_n & C_n \\
\Theta_{-, +}^{(n) \, \dagger} & - C_n & S_n & 1 & 0 \\
\Theta_{-, -}^{(n) \, \dagger} & S_n & C_n & 0 & 1 
\end{array}
\end{eqnarray}
gives the following matrix elements
\begin{eqnarray}
\Theta_{\pm_\kappa \pm_m}^{(n) \, \dagger} \gamma^{\theta_n}_n \Theta^{(n)}_{\pm_\kappa \pm_m} &=& 0 \, , \nonumber \\
\Theta_{\pm_\kappa \pm_m}^{(n) \, \dagger} \gamma^{\phi_n}_n \Theta^{(n)}_{\pm_\kappa \pm_m} &=& - \pm_\kappa \pm_m \mathrm i S_n \, , \nonumber \\
\Theta_{\pm_\kappa \pm_m}^{(n) \, \dagger} \gamma^{\mathcal K_n}_n \Theta^{(n)}_{\pm_\kappa \pm_m} &=& \pm_m \mathrm i C_n \, , \nonumber \\
\Theta_{\pm_\kappa \pm_m}^{(n) \, \dagger} \partial_{\theta_n} \Theta^{(n)}_{\pm_\kappa \pm_m} &=& 0 \, , \nonumber \\
\Theta_{\pm_\kappa \pm_m}^{(n) \, \dagger} \gamma^{\phi_n}_n \partial_{\theta_n} \Theta^{(n)}_{\pm_\kappa \pm_m} &=& - \pm_\kappa \pm_m \mathrm i C_n \, /2 \, .
\end{eqnarray}
On the other hand, the covariant derivatives are explicitly given by
\begin{eqnarray}
\nabla_{\theta_k}^{(n)} &=& \left[ \prod\limits_{k<j\le n}^{j \equiv n \, \mathrm{mod} \, 2} \frac{1}{C_j} \right] \partial_{\theta_k} \nonumber \\
&&- \frac{\mathrm i}{2} \sum\limits_{k < j \le n}^{j \equiv n \, \mathrm{mod} \, 2} \left[ \prod\limits_{j<l\le n}^{l \equiv n \, \mathrm{mod} \, 2} \frac{1}{C_l} \right] \frac{S_j}{C_j} \, \gamma_j^{\phi_j} \gamma_{j-2}^{\theta_k} \, , \nonumber \\
\nabla_{\phi_k}^{(n)} &=& \left[ \prod\limits_{k<j\le n}^{j \equiv n \, \mathrm{mod} \, 2} \frac{1}{C_j} \right] \left[ \frac{1}{S_k} \partial_{\phi_k} - \frac{C_k}{2 S_k} \gamma_k^{\mathcal K_k} \right] \nonumber \\
&&- \frac{\mathrm i}{2} \sum\limits_{k < j \le n}^{j \equiv n \, \mathrm{mod} \, 2} \left[ \prod\limits_{j<l\le n}^{l \equiv n \, \mathrm{mod} \, 2} \frac{1}{C_l} \right] \frac{S_j}{C_j} \, \gamma_j^{\phi_j} \gamma_{j-2}^{\phi_k} \, . 
\end{eqnarray}
In the case of $n$ being odd we have
\begin{eqnarray}
\gamma_1^{\phi_1} &=& \mathrm i \gamma_3^{\mathcal K_3} \gamma_3^{\phi_1} = - \mathrm i \gamma_3^{\theta_3} \gamma_3^{\phi_3} \gamma_3^{\phi_1} \equiv - \mathrm i \, , \nonumber \\
\theta_1 &\equiv& \pi / 2 \; , \; C_1 \equiv 0 \; , \; S_1 \equiv 1 \, .
\end{eqnarray}
For the construction of the spherically symmetric stress-energy-tensor we now need the expectation value of $\nabla^{(n)}_k$ and $\gamma_n^k \nabla^{(n)}_j$ with $\Theta_{\epsilon_\kappa, \boldsymbol{\epsilon}}$. Write these as $\langle  \Gamma \rangle = \Theta^\dagger_{\epsilon_\kappa, \boldsymbol{\epsilon}} \Gamma \Theta_{\epsilon_\kappa, \boldsymbol{\epsilon}}$, for the matrix element of $\Gamma$. 
The following identity proves to be useful,
\begin{eqnarray}
\gamma^{\mathcal \alpha_k}_n &=& \left[ \prod\limits_{k < j \le n}^{j \equiv n \, \mathrm{mod} \, 2} \mathrm i \gamma_j^{\mathcal K_j} \right] \gamma^{\alpha_k}_k \, ,  
\end{eqnarray}
with $j \equiv k \equiv n \, \mathrm{mod} \, 2$, $n \ge k < j$ and $\alpha \in \{\theta, \phi, \mathcal K\}$,

After some tedious algebra we find the following expressions
\begin{eqnarray}
\langle \nabla_{\theta_k}^{(n)} \rangle &=& 0 = \langle \gamma_n^{\theta_k} \rangle \, , \nonumber \\
\langle \nabla_{\phi_k}^{(n)}\rangle &=& \left\{ \frac{1}{\pi_1^{(n)}} - \Sigma_{k, n} \right\} \frac{\mathrm i \epsilon_k}{2} S_k \, , \nonumber \\
\langle \gamma_n^{\phi_k} \rangle &=& - \mathrm i \epsilon_k \epsilon_\kappa \pi_k^{(n)} S_k \, , \nonumber \\
\langle \gamma_n^{\theta_j} \nabla^{(n)}_{\theta_k} \rangle &=& \frac{\epsilon_\kappa}{2} \delta_{j k} \, , \nonumber \\
\langle \gamma_n^{\phi_j} \nabla^{(n)}_{\theta_k} \rangle &=& -\frac{\mathrm i}{2} \frac{\delta_k}{\pi_k^{(n)}} \epsilon_\kappa C_k \, \delta_{j k} \, , \nonumber \\
\langle \gamma_n^{\theta_j} \nabla^{(n)}_{\phi_k} \rangle &=& - \frac{\mathrm i}{2} S_k \pi_k^j S_j \epsilon_\kappa \epsilon_k \, , \, j > k \, ,\nonumber \\
\langle \gamma_n^{\theta_k} \nabla^{(n)}_{\phi_k} \rangle &=& \frac{i C_k}{2} \epsilon_\kappa \epsilon_k \, , \nonumber \\
\langle \gamma_n^{\phi_j} \nabla^{(n)}_{\phi_k} \rangle &=& \left\{ \frac{1}{\pi_k^{(j)}} + \sum\limits_{k < l < j}^{l \equiv n \, \mathrm{mod} \, 2} \pi_k^l \frac{S_l^2}{C_l} - \pi_k^j \frac{1}{C_j} \right\} \nonumber \\
&&\times \frac{\epsilon_\kappa \epsilon_k \epsilon_j}{2} S_j S_k \;\;\;\;\; \;\;\;\;\; \;\;\;\;\; \;\;\;\;\; \;\;\;\;\; \;\;\;\;\; \;\; , \, j > k \, , \nonumber \\
\langle \gamma_n^{\phi_k} \nabla_{\phi_k}^{(n)} \rangle &=& \frac{\epsilon_\kappa}{2} \, , \nonumber \\
\langle \gamma_n^{\phi_j} \nabla^{(n)}_{\phi_k} \rangle &=& 0 = \langle \gamma_n^{\theta_j} \nabla_{\phi_k}^{(n)} \rangle \, , \, j < k \, , \nonumber \\
\langle \gamma^{\mathcal K_k}_n \rangle &=& - \mathrm i \delta_k \pi_k^{(n)} C_k \, ,
\end{eqnarray}
where we have defined
\begin{eqnarray}
\Sigma_{k, m} &=& \sum\limits_{k < j \le m}^{j \equiv n \, \mathrm{mod} \, 2} \pi_k^j \frac{S^2_j}{C_j} \frac{1}{\pi_j^{(m)}} \, , \nonumber \\
\delta_k &=& \left[\prod\limits_{k \le l \le n}^{l \equiv n \, \mathrm{mod} \, 2} - \epsilon_l \right] \, , \nonumber \\
\pi_k^j &=& \prod\limits_{k < l < j}^{l \equiv n \, \mathrm{mod} \, 2} C_l \, , \nonumber \\
\pi_k^{(j)} &=& \prod\limits_{k < l \le j}^{l \equiv n \, \mathrm{mod} \, 2} C_l \, .
\end{eqnarray}
It is important to note that
\begin{eqnarray}
\sum_{\boldsymbol \epsilon} \langle \nabla_{\theta_k}^{(n)} \rangle &=& \sum_{\boldsymbol \epsilon} \langle \nabla_{\phi_k}^{(n)} \rangle = \sum_{\boldsymbol \epsilon} \langle \gamma_n^{\phi_j} \nabla_{\theta_k}^{(n)} \rangle = \sum_{\boldsymbol \epsilon} \langle \gamma_n^{\theta_l} \nabla_{\theta_k}^{(n)} \rangle \nonumber \\
&=& \sum_{\boldsymbol \epsilon} \langle \gamma_n^{\theta_j} \nabla_{\phi_k}^{(n)} \rangle = \sum_{\boldsymbol \epsilon} \langle \gamma_n^{\phi_l} \nabla_{\phi_k}^{(n)} \rangle = \sum_{\boldsymbol \epsilon} \langle \gamma_n^{\theta_k} \rangle \nonumber \\
&=& \sum_{\boldsymbol \epsilon} \langle \gamma_n^{\phi_k} \rangle = \sum_{\boldsymbol \epsilon} \langle \gamma^{\mathcal K_k}_n \rangle = 0 \, ,
\end{eqnarray}
for $l \neq k$ and the sum is over all possible sign vectors $\boldsymbol{\epsilon}$. 
Hence the non-diagonal terms sum to zero.

The non-vanishing sums are in the diagonal parts, which result in
\begin{eqnarray}
\sum_{\boldsymbol \epsilon} \langle \gamma_n^{\theta_k} \nabla^{(n)}_{\theta_k} \rangle = \sum_{\boldsymbol \epsilon} \langle \gamma_n^{\phi_k} \nabla_{\phi_k}^{(n)} \rangle = 2^{\left\lfloor \frac{n}{2} \right\rfloor} \frac{\epsilon_\kappa}{2} \, .
\end{eqnarray}

\subsection{Combining the spinors}

We will use the above expressions to construct a field configuration with a spherically symmetric energy momentum tensor.

Fix a sign $\epsilon_\kappa$ for a lowest angular eigenvalue $\kappa = \epsilon_\kappa \frac{d-2}{2}$ of the $(d-2)$-sphere. Define $2^{\lfloor \frac{d-2}{2} \rfloor}$ spinor fields as in equation (\ref{eqn:spinor1}), 
\begin{eqnarray}
\Psi_{\boldsymbol{\epsilon}} = \mathrm e^{-\mathrm i \omega t} \phi_\kappa \otimes \Theta_{\kappa, \boldsymbol \epsilon} 
\end{eqnarray}
but labeling explicitly all the  allowed sign combinations of $\boldsymbol{\epsilon}$.
We then combine these spinors in an incoherent superposition so that
\begin{eqnarray}\label{eqn:ansatz}
\Psi_\text{conf.} : \!\! &=& \bigoplus_{\boldsymbol{\epsilon}} \Psi_{\boldsymbol{\epsilon}} = \bigoplus_{\boldsymbol{\epsilon}} \mathrm e^{-\mathrm i \omega t} \phi_\kappa \otimes \Theta_{\kappa, \boldsymbol \epsilon} \nonumber \\ 
&=& \mathrm e^{-\mathrm i \omega t} \phi_\kappa \otimes \bigoplus_{\boldsymbol{\epsilon}} \Theta_{\kappa, \boldsymbol \epsilon}
\end{eqnarray}
written here as a formal sum ranging over all possible values for $\boldsymbol{\epsilon}$. Note in the last step of equation (\ref{eqn:ansatz}) we have made use of the fact that the radial and temporal dependence of each individual spinor is the same for all of them. 

In the spacetime of the metric given by equation (\ref{eqn:spherically_spacetime})
the covariant derivatives are explicitly given by
\begin{eqnarray}
\nabla_t &=& \frac{1}{\sqrt{N} \, \sigma} \partial_t + \sqrt{N} \frac{\mathrm d \ln \sqrt{\sqrt{N} \, \sigma}}{\mathrm d r} \gamma^t \gamma^r \, , \nonumber \\
\nabla_r &=& \sqrt{N} \partial_r \, , \nonumber \\
\nabla_j &=& \frac{1}{r} \nabla^{(d-2)}_j + \frac{\sqrt{N}}{2 r} \gamma^t \gamma^j_{d-2} \, , 
\end{eqnarray}
with $j$ being an index on the $(d-2)$-sphere and
\begin{eqnarray}
\nabla_j^{(d-2)} = \mathbf e_j^{d-2} + \frac{1}{2} \Gamma^{d-2}_{|k l|} (\mathbf e^{d-2}_j) \gamma_{d-2}^k \gamma_{d-2}^l
\end{eqnarray}
being the covariant derivative on the $(d-2)$-sphere and $\gamma_{d-2}^j := \gamma^t \gamma^r \gamma^j$ being the $\gamma$-matrices of the $(d-2)$-sphere.

Using for the sphere the same vielbein as in the previous sections, and after some algebraic manipulations in which one needs to make use of the expressions we have derived in section \ref{s-e-tensor_components}, one arrives at the following energy momentum tensor (note that it is written in vielbein components)
\begin{eqnarray}
T_{tt} &=& - 2^{\left\lfloor \frac{d-2}{2} \right\rfloor+1} \frac{\Re(\omega)}{\sqrt{N} \, \sigma} \phi^\dagger_\kappa \phi_\kappa \, , \nonumber \\
T_{tr} &=& \frac{2^{\left\lfloor \frac{d-2}{2} \right\rfloor + 1}}{\sqrt{N} \, \sigma} \Re(\omega) \phi^\dagger_\kappa \gamma^t \gamma^r \phi_\kappa \, , \nonumber \\
T_{rr} &=&- 2^{\left\lfloor \frac{d-2}{2} \right\rfloor + 1}  \sqrt{N}\, \Im ( \phi^\dagger_\kappa \gamma^t \gamma^r \partial_r \phi_\kappa) \, \nonumber \\
T_{t j} &=& 0 = T_{r j} \, , \nonumber \\
T_{j k} &=& - \frac{\epsilon_\kappa}{r} 2^{\left\lfloor \frac{d-2}{2} \right\rfloor} \Im ( \phi^\dagger_\kappa \gamma^r \phi_\kappa ) \delta_{jk} \, , 
\end{eqnarray}
where we have used the radial equation for $\phi_\kappa$ to simplify some expressions,
\begin{eqnarray}\label{eqn:phi_equation}
&&\left[ \frac{\omega}{\sqrt{N} \sigma} \gamma^t + \mathrm i \sqrt{N} \gamma^r \left( \frac{\mathrm d}{\mathrm d r} + \frac{\mathrm d}{\mathrm d r} \ln \sqrt{ \sqrt{N} \, \sigma r^{d-2}} \right) \right. \nonumber \\
&&\left. + \frac{\mathrm i}{r} \gamma^t \gamma^r \kappa - m \right] \phi_\kappa = 0 \, .
\end{eqnarray}
One can easily see that this tensor is diagonal on the spatial components and thus spherically symmetric.

Note however that the tensor has in general a non-trivial $t-r$ component. This means the configuration in general has a radial flux, and will force the configuration to be time dependent. If we want to obtain solutions compatible with the  static metric (\ref{eqn:spherically_spacetime}), we have to require the radial current to vanish everywhere, meaning $\phi^\dagger_\kappa \gamma^t \gamma^r \phi_\kappa = 0$ (no-flux).

The only thing left is to choose a particular representation of the remaining $\gamma$-matrices and spinor components,
\begin{eqnarray}
\gamma^t = \left[ \begin{array}{cc} 0 & 1 \\ 1 & 0 \end{array} \right] \; , \; \gamma^r = \left[ \begin{array}{cc} 0 & -1 \\ 1 & 0 \end{array} \right] \; , \; \phi_\kappa = \left[ \begin{array}{c} \phi_1 \\ \phi_2 \end{array} \right] \, .
\end{eqnarray}
The no-flux condition reads
\begin{eqnarray}
|\phi_1|^2 = |\phi_2|^2 \, .
\end{eqnarray}
The following parametrization incorporates the no-flux condition
\begin{eqnarray}\label{eqn:r-original}
\phi_1 = 2^{-\frac{1}{2}\left\lfloor \frac{d-2}{2} \right\rfloor} \, \hat \phi \; , \; \phi_2 = 2^{-\frac{1}{2}\left\lfloor \frac{d-2}{2} \right\rfloor} \, \mathrm{e}^{\mathrm i \nu} \hat \phi \, , 
\end{eqnarray}
with $\hat \phi$ a complex function and $\nu$ a real valued function. 

Using this Ansatz and representation in the equation (\ref{eqn:phi_equation}) and after some algebra we get the following non-linear first order system of differential equations for $\hat \phi$ and $\nu$
\begin{eqnarray}\label{eqn:diff-equations}
\frac{\mathrm d \ln |\hat \phi|}{\mathrm d r} &=&\frac{1}{\sqrt{N}} \Im \left\{ \left(m + \frac{\mathrm i \kappa}{r} \right) \mathrm e^{\mathrm i \nu} \right\} - \frac{\mathrm d \ln \sqrt{\sqrt{N} \, \sigma \, r^{d-2}}}{\mathrm d r} \, , \nonumber \\
\frac{\mathrm d \nu}{\mathrm d r} &=& \frac{2}{\sqrt{N}} \Re \left\{ \left( m + \frac{\mathrm i \kappa}{r} \right) \mathrm e^{\mathrm i \nu} \right\} - \frac{2 \omega}{N \sigma} \, .
\end{eqnarray}
The equation for $\nu$ forces the frequency $\omega$ to be real. Note that the phase of $\hat \phi$ does not vary with $r$ and is not a dynamical quantity.

The stress-energy tensor in the vielbein components simplifies into
\begin{eqnarray}
T_{tt} &=& - \frac{4 \omega}{\sqrt{N} \, \sigma} |\hat \phi|^2 \, , \nonumber \\
T_{rr} &=& 4 \left\{ m \cos \nu  - \frac{\kappa \sin \nu}{r} - \frac{\omega}{\sqrt{N} \, \sigma} \right\} |\hat \phi|^2  \nonumber \\
&=& 2 \sqrt{N} \,  \frac{\mathrm d \nu}{\mathrm d r} |\hat \phi|^2 \, , \nonumber \\
T_{t j} &=& 0 = T_{r j} = T_{t r} \, , \nonumber \\
T_{j k} &=& \frac{2 \epsilon_\kappa \sin \nu}{r} |\hat \phi|^2  \delta_{jk} \, .
\end{eqnarray}

An important quantity we can calculate is the time component of the net current in the vielbein, the Dirac density:
\begin{eqnarray}
j^0_\text{net} = \sum\limits_{\boldsymbol \epsilon} \phi^\dagger_\kappa \phi_\kappa = 2 |\hat \phi|^2 \, .
\end{eqnarray}
We can see that all the components of the stress-energy tensor are proportional to the Dirac density $j^0_\text{net}$.

\subsection{A comment on the time-dependent case}

Although we are mainly interested in static metrics, the previous Ansatz can be easily generalized to accommodate the time-dependent case.

In this case the metric functions $\sigma$ and $N$ also have to depend on time. But it is also necessary to change the Ansatz for the $\Psi_{\boldsymbol \epsilon}$ to
\begin{eqnarray}
\Psi_{\boldsymbol \epsilon} = \phi_\kappa(t,r) \otimes \Theta_{\kappa, \boldsymbol \epsilon} \, ,
\end{eqnarray}
meaning there is no harmonic time dependence in the fields.

With these changes, the stress-energy tensor becomes
\begin{eqnarray}
T_{tt} &=& \frac{2^{\left\lfloor \frac{d-2}{2} \right\rfloor +1}}{\sqrt{N} \, \sigma}  \Im ( \phi_\kappa^\dagger \partial_t \phi_\kappa) \, , \nonumber \\
T_{tr} &=& - \frac{2^{\left\lfloor \frac{d-2}{2} \right\rfloor + 1}}{\sqrt{N} \, \sigma} \Im (\phi_\kappa^\dagger \gamma^t \gamma^r \partial_t \phi_\kappa) \, , \nonumber \\
T_{rr} &=& - 2^{\left\lfloor \frac{d-2}{2} \right\rfloor + 1} \sqrt{N}  \, \Im ( \phi_\kappa^\dagger \gamma^t \gamma^r \partial_r \phi_\kappa ) \, , \nonumber \\
T_{tj} &=& 0 = T_{r j} \, , \nonumber \\
T_{j k} &=& - \frac{\epsilon_\kappa}{r} 2^{\left\lfloor \frac{d-2}{2} \right\rfloor} \Im ( \phi^\dagger_\kappa \gamma^r \phi_\kappa ) \delta_{jk} \, .
\end{eqnarray}
The equation fulfilled by $\phi_\kappa$ is now a partial differential equation,
\begin{eqnarray}
&&\left\{ \frac{\mathrm i}{\sqrt{N} \, \sigma} \gamma^t \left[ \partial_t - \partial_t \ln N^{1/4} \right] \right.  \\
&&\left. + \mathrm i \sqrt{N} \gamma^r \left[ \partial_r + \partial_r \ln \sqrt{\sqrt{N} \sigma r^{d-2}} \;\right] + \frac{\mathrm i \kappa}{r} \gamma^t \gamma^r - m \right\} \phi_\kappa = 0 \, . \nonumber
\end{eqnarray}

In the following we will consider only a static space-time, and assume the Dirac fields possess a harmonic time-dependence.

\section{Effective action}
\label{section_action}

With the construction we have developed in the previous section, it is possible to simplify the part of the action (\ref{eqn:action_Diracs}) containing the collection of Dirac fields, 
\begin{eqnarray}
S &=& \int \mathrm d x^4 \sqrt{|g|} \left(R + \frac{2 \alpha_g}{\sqrt{|g|}} \mathcal L_\text{spinor} + \Lambda \right)  \\
&=& A_{d-2} \int \mathrm d t \int \mathrm d r \sigma r^{d-2} \nonumber \\
&& \times \left\{ \left[ N^{\prime \prime} + \frac{3 N^\prime \sigma^\prime}{\sigma} + \frac{2 N}{\sigma} \sigma^{\prime \prime} + \frac{2(d-2) N^\prime}{r} \right. \right. \nonumber \\ &&\left. \left.  + \frac{2 (d-2) N}{\sigma r} \sigma^\prime + \frac{(d-2)(d-3)(N-1)}{r^2} \right] + \Lambda \right. \nonumber \\
&& \left. + 2 \alpha_g \left[ \frac{\sqrt{N}}{2} \frac{\mathrm d \nu}{\mathrm d r} - \Re \left\{ \left(m + \frac{\mathrm i \kappa}{r} \right) \mathrm e^{\mathrm i \nu} \right\} +  \frac{\omega}{\sqrt{N} \, \sigma}  \right]  | \hat \phi |^2 \right\}  \nonumber
\\
&& \equiv A_{d-2} \int \mathrm d t \int \mathrm d r
\mathcal L_\text{eff}  \nonumber
 \, ,
\end{eqnarray}
where we have defined the effective Lagrangian $\mathcal L_\text{eff}$. 
With this the equations of motion read, using $j_\text{net}^0 = 2 |\hat \phi|^2$,
\begin{eqnarray}
N^\prime &=& - \frac{d-3}{r} (N-1) \nonumber \\
&&- \frac{\alpha_g r}{d-2} \left[ \frac{\sqrt{N}}{2} \nu^\prime - \Re \left\{\left(m + \frac{\mathrm i \kappa}{r} \right) \mathrm e^{\mathrm i \nu} \right\} \right] j^0_\text{net}  \nonumber \\
&&- \Lambda r\, , \nonumber \\
\sigma^\prime &=& \frac{\alpha_g}{2(d-2)} \frac{r}{\sqrt{N}} \left[ \frac{\sigma}{2} \nu^\prime - \frac{\omega}{N} \right] j^0_\text{net} \, , \nonumber \\
\nu^\prime &=& \frac{2}{\sqrt{N}} \Re\left\{ \left(m + \frac{\mathrm i \kappa}{r} \right) \mathrm e^{\mathrm i \nu} \right\} - \frac{2 \omega}{N \sigma} \, ,  \\
(\ln j^0_\text{net})^\prime &=& \frac{2}{\sqrt{N}} \Im \left\{ \left(m + \frac{\mathrm i \kappa}{r} \right) \mathrm e^{\mathrm i \nu} \right\} - ( \ln [ \sqrt{N} \sigma r^{d-2} ] )^\prime \, . \nonumber 
\end{eqnarray}
or after using the equation for $\nu^\prime$ in the equation for $N^\prime$ and defining 
\begin{eqnarray}
\mathrm e^{\lambda(r)} = \sqrt{N} \sigma r^{d-2} j^0_\text{net} \, ,
\nonumber \\
\xi(r) = \frac{2}{\sqrt{N}} \left(m + \frac{\mathrm i \kappa}{r} \right) \mathrm e^{\mathrm i \nu} \, ,
\end{eqnarray}
we have
\begin{eqnarray}\label{eqn:eq_of_motion_2}
N^\prime &=& \frac{d-3}{r} (1-N) + \frac{\alpha_g \omega}{d-2} \frac{1}{N \sigma^2 r^{d-3}} \mathrm e^\lambda -  \Lambda r\, , \nonumber \\
\sigma^\prime &=& \frac{\alpha_g}{2(d-2)} \frac{1}{N r^{d-3}} \left[ \frac{1}{2} \Re (\xi) - \frac{2 \omega}{N \sigma} \right] \mathrm e^\lambda \, , \nonumber \\
\nu^\prime &=& \Re ( \xi ) - \frac{2 \omega}{N \sigma} \, , \nonumber \\
\lambda^\prime &=& \Im ( \xi )\, .
\end{eqnarray}

Another useful way to write this is using
\begin{eqnarray}
\hat \phi \mathrm e^{\mathrm i \nu / 2} = g - \mathrm i f \, .
\end{eqnarray}
This means that
\begin{eqnarray}\label{eqn:set}
j_\text{net}^0 &=&2|\hat \phi|^2 = 2(f^2+g^2) \, \, , 
 \, \,  \, \, 
\mathrm e^{\mathrm i \nu} = \frac{g- \mathrm i f}{g+\mathrm i f} \, \, , \nonumber \\
\cos \nu &=& \frac{g^2 - f^2}{f^2 + g^2} \, \, , 
 \, \,  \, \, 
\sin \nu = - \frac{2 f g}{f^2 + g^2} \, .
\end{eqnarray}
With this the effective Lagrangian for the spinor part is especially simple
\begin{eqnarray}
\mathcal L_\text{spinor} &=& r^{d-2} \sigma \sqrt{N} \left( f \frac{\mathrm d g}{\mathrm d r} - g \frac{\mathrm d f}{\mathrm d r} \right) \nonumber \\
&&- r^{d-2} \sigma \left( m [ g^2 - f^2] + \frac{2 \kappa}{r} f g \right) \nonumber \\
&&+ \frac{r^{d-2} \omega}{\sqrt{N}} (f^2 + g^2) \, .
\end{eqnarray}
The equations of motion are 
\begin{eqnarray}
N^\prime &=& - \frac{d-3}{r} (N-1) + \frac{\alpha_g r}{d-2} \frac{2 \omega}{\sqrt{N} \sigma} (f^2 + g^2) -\Lambda r \, , \nonumber \\
\sigma^\prime &=& \frac{\alpha_g}{d-2} \frac{r}{N} \left\{ m \sigma (g^2 - f^2) + \frac{2 \kappa \sigma}{r} f g - \frac{2 \omega}{\sqrt{N}} (f^2 + g^2) \right\} \nonumber \\
f^\prime &=& - \left\{  \frac{\kappa}{\sqrt{N} r} + \frac{\mathrm d \ln \sqrt{r^{d-2} \sigma \sqrt{N}}}{\mathrm d r} \right\} f 
- \left\{ \frac{m}{\sqrt{N}} - \frac{\omega}{N \sigma} \right\} g \, , \nonumber \\
g^\prime &=& \left\{  \frac{\kappa}{\sqrt{N} r} - \frac{\mathrm d \ln \sqrt{r^{d-2} \sigma \sqrt{N}}}{\mathrm d r} \right\} g 
- \left\{ \frac{m}{\sqrt{N}} + \frac{\omega}{N \sigma} \right\} f \, . \nonumber \\
\end{eqnarray}
This form is useful for numerical calculations \cite{Blazquez-Salcedo:2019qrz}.

Finally, if we assume that $\sigma>0$, which we can always do without loss of generality, there is a convenient way to redefine the spinor functions by setting
\begin{eqnarray}
f & = & \sqrt{ \frac{d-2}{\alpha_g r^{d-2} \sigma \sqrt{N}}} \hat{f} \, , \nonumber \\
g & = & \sqrt{ \frac{d-2}{\alpha_g r^{d-2} \sigma \sqrt{N}}} \hat{g} \, .
\end{eqnarray}
This is convenient, because it makes the form of the field equations a bit more compact,
\begin{eqnarray}\label{eqn:diff_f_g}
N^\prime &=& - \frac{d-3}{r} (N-1) + \frac{2 \omega}{N \sigma^2 r^{d-3}} (\hat{f}^2 + \hat{g}^2) - \Lambda r \, , \nonumber \\
\sigma^\prime &=& \frac{1}{N^{3/2} r^{d-3}} \left\{ m (\hat{g}^2 - \hat{f}^2) + \frac{2 \kappa}{r} \hat{f} \hat{g} - \frac{2 \omega}{\sigma \sqrt{N}} (\hat{f}^2 + \hat{g}^2) \right\} \nonumber \\
\hat{f}^\prime &=& - \frac{\kappa}{\sqrt{N} r} \hat{f} - \left\{\frac{m}{\sqrt{N}} - \frac{\omega}{N \sigma} \right\} \hat{g} \, , \nonumber \\
\hat{g}^\prime &=&   \frac{\kappa}{\sqrt{N} r}  \hat{g} - \left\{ \frac{m}{\sqrt{N}} + \frac{\omega}{N \sigma} \right\} \hat{f} \, .
\end{eqnarray}
This will be helpful in the next sections. Let us note here that with these definitions the Dirac density is
\begin{eqnarray}
j_\text{net}^0 = \frac{2 (d-2)}{\alpha_g r^{d-2} \sigma \sqrt{N}} \left[ \hat{f}^2 + \hat{g}^2 \right] \, .
\end{eqnarray}

\section{Analytical solutions}
\label{section_sols}

Of course a fundamental question that immediately arises is if, for some set of parameters, the previous system of equations possesses physically meaningful configurations, and what is their interpretation. In the following, we will focus on cases without cosmological constant ($\Lambda=0$).

 Soliton-like solutions of this system in several dimensions have been presented in \cite{Blazquez-Salcedo:2019qrz}. These solutions, to our knowledge, can only be constructed numerically. The solutions (sometimes called Dirac stars, although they are not expected to have any connection with realistic astrophysical objects) are regular everywhere, and share many features with similar self-gravitating soliton-like configurations found with massive bosonic fields.

In this section we want to present several analytical solutions that the previous system possesses. We will analyze in detail the physical and geometrical properties of these solutions.

\subsection{Multi-Dirac wormhole with unbounded spinors}
\label{section_wormhole}

\subsubsection{The solution}

Let us specialize to a massless $(m=0)$ field which does not vary in time $(\omega = 0)$. The differential equations (\ref{eqn:diff_f_g}) simplify in this case to
\begin{eqnarray}
	N^\prime &=& - \frac{d-3}{r} (N-1)  \, , \nonumber \\
	\sigma^\prime &=& \frac{1}{N^{3/2} r^{d-3}}  \frac{2 \kappa}{r} \hat{f} \hat{g}  \, , \nonumber \\
	\hat{f}^\prime &=& - \frac{\kappa}{\sqrt{N} r} \hat{f} \, , \nonumber \\
	\hat{g}^\prime &=&   \frac{\kappa}{\sqrt{N} r}  \hat{g}\, .
	\label{wh_eqs}
\end{eqnarray}
The differential equation for $N$ means that $N = 1 - (\mu / r)^{d-3}$ with $\mu$ being a constant. 
Without loss of generality we can fix positive angular momentum of the fields, $\epsilon_\kappa=1$, $\kappa=\frac{d-2}{2}$. The solutions for $f$ and $g$ are
\begin{eqnarray}
\label{sol_wh}
	\hat{f} &=& -\exp \left( (c_0+c_\Delta)/2 - \frac{2 \kappa}{d-3} \mathrm{artanh} \sqrt{N} \right) \, , \nonumber \\
	\hat{g} &=& \exp \left( (c_0-c_\Delta)/2 + \frac{2 \kappa}{d-3} \mathrm{artanh} \sqrt{N} \right) \, ,
\end{eqnarray}
with $c_0, c_\Delta \in \mathbb R$. Notice that $\hat f \hat g = -\mathrm e^{c_0} \in \mathbb R_{\le 0}$. This simplifies the differential equation for $\sigma$, which now reads
\begin{eqnarray}
	\sigma^\prime &=& - 2 \kappa \mathrm e^{c_0} \frac{1}{N^{3/2} r^{d-2}}  
	=\frac{4 \kappa \mathrm e^{c_0}}{\mu^{d-3} (d-3)} \frac{\mathrm d}{\mathrm d r} \frac{1}{\sqrt{N}} \, .
\end{eqnarray}
This is easily integrated to be
\begin{eqnarray}
	\label{eqn:sigma1}
	\sigma(r) = c_\sigma +  \frac{4 \kappa \mathrm e^{c_0}}{\mu^{d-3} (d-3)}\frac{1}{\sqrt{N}} 
	\, ,
\end{eqnarray}
with $c_\sigma \in \mathbb R$. 
In total, the solution is parameterized by three real constants $c_\sigma$, $c_0$ and $c_\Delta$, in addition to $\mu$, the coupling constant $\alpha_g$ and the dimension $d$. 
However the parameters satisfy several relations. 

\subsubsection{Constraints on the solution}

Our first requirement is for $\sigma$ to be a positive function in all of the domain $r \in [\mu,\infty)$. An analysis of equation (\ref{eqn:sigma1}) reveals that 
this is only possible if $c_\sigma > - \frac{2 \mathrm e^{c_0}}{\mu^{d-3}} \frac{d-2}{d-3}$

The second requirement is to reach the standard Minkowski metric at infinity. This means that
\begin{eqnarray}
	g_{tt}=N \sigma^2 \rightarrow 1 \; , \; \text{for} \; r \rightarrow \infty \, .
\end{eqnarray}
Taking into account the previous condition for $\sigma>0$, this implies the following relation for $c_0$
\begin{eqnarray}
	\label{eqn:csigma}
	e^{c_0}=(1-c_\sigma)\frac{\mu^{d-3}(d-3)}{4\kappa}
	\, .
\end{eqnarray}%
Hence we can write
\begin{eqnarray}
	\label{eqn:sigma2}
	\sigma(r) = c_\sigma+ \frac{1-c_\sigma}{\sqrt{N}} 
	\, ,
\end{eqnarray}
and it is easy to see that the $g_{tt}$ component behaves asymptotically like
\begin{eqnarray}
\label{eq_gtt_asymp}
	g_{tt} = 1 - \left(\frac{\mu}{r}\right)^{d-3}c_{\sigma} + O(1/r^{d-2}) \, .
\end{eqnarray}
Hence we can clearly see that the parameter $c_{\sigma}$ is related with the mass of the solution.

\subsubsection{Massless wormhole}

Let us explore the physical meaning of this metric. To simplify the discussion, let us first look at the particular case with $c_\sigma = 0$. Thus $\sigma(r) = 1 / \sqrt{N}$. The metric is very simple,
\begin{eqnarray}
	\mathrm d s^2 = \mathrm d t^2 - N^{-1}(r) \mathrm d r^2 - r^2 \mathrm d \Omega^2_{d-2} \, ,
\end{eqnarray}
with $N = 1 - (\mu/r)^{d-3}$. This looks like the metric of a traversable wormhole \cite{Visser:1995cc}. Let us make the following coordinate transformation in this metric, $\rho = \sqrt{N}$. This leads to
\begin{eqnarray}
	\label{wormhole_rho}
	\mathrm d s^2 &=& \mathrm d t^2 - \frac{4}{\mu^{2(d-3)} (d-3)^2} \left[ \frac{\mu^{d-3}}{1 - \rho^2} \right]^\frac{2(d-2)}{d-3} \mathrm d \rho^2 \nonumber \\
	&&- \left[ \frac{\mu^{d-3}}{1 - \rho^2} \right]^\frac{2}{d-3} \mathrm d \Omega_{d-2}^2 \, .
\end{eqnarray}
For $r \in [\mu, \infty )$ we have $\rho \in [0, 1)$ mapped such that $r = \infty \mapsto 1 = \rho$. 
In this coordinate system, it is possible to extend the above metric (\ref{wormhole_rho}) to $\rho \in (-1, 1)$. Thus the above metric corresponds to a wormhole connecting two asymptotic regions at $\rho = \pm 1$. The sphere with minimal surface has radius $r = \mu$, which corresponds to the throat of the wormhole as we will explicitly see later. Note that the Ricci scalar of this metric vanishes, but the Krestchmann scalar is finite. For $d=4$, $K=6(1-\rho^2)^6/\mu^4$; for $d=5$, $K=24(1-\rho^2)^4/\mu^4$, etc...

An interesting property of the above geometry is that the temporal part of the metric is essentially not curved and thus a test mass can rest at a fixed radius $\rho$ without moving. %

We can compute the mass using the standard Komar integral 
\begin{eqnarray}
M = - \frac{1}{16 \pi} \frac{d-2}{d-3} \int\limits_{r \rightarrow \infty} \ast \mathbf d \xi_t =0 \, ,
\end{eqnarray}
with $\xi_t$ being the one form dual to the Killing vector $K_t = \partial_t$ and the integral being over a $(d-2)$-sphere at infinity. For the metric (\ref{wormhole_rho}) this is trivially zero.
Hence the mass of this wormhole vanishes, as it was expected from the form of the asymptotical relation (\ref{eq_gtt_asymp}).

\subsubsection{Massive wormhole}

Now let us explore the more general case with $c_\sigma \neq 0$. 
It is convenient to change again the radial coordinate to $\rho = \sqrt{N}$, similar to what we have seen for $c_{\sigma}=0$, because we can extend its range  from $\rho \in [0, 1)$ to $\rho \in (-1, 1)$. We have a metric slightly different than the one in the previous case,
\begin{eqnarray}
	\label{wormhole_rho_2}
	\mathrm d s^2 &=& \left[ 1 - c_\sigma \left( 1 - \rho \right) \right]^2 \mathrm d t^2 \nonumber \\
	&&- \frac{4}{\mu^{2(d-3)} (d-3)^2} \left[ \frac{\mu^{d-3}}{1 - \rho^2} \right]^\frac{2(d-2)}{d-3} \mathrm d \rho^2 \nonumber \\
	&&- \left[ \frac{\mu^{d-3}}{1 - \rho^2} \right]^\frac{2}{d-3} \mathrm d \Omega_{d-2}^2 \, .
\end{eqnarray}
Note that with $c_{\sigma}\neq0$, the $g_{tt}$ component depends on $\rho$. The asymptotical behaviour at $ \rho=1$ (i.e., $r\to \infty$) is given by equation (\ref{eq_gtt_asymp}), since
\begin{eqnarray}
\rho = \sqrt{ 1 - \left( \frac{\mu}{r} \right)^{d-3} } = 1 - \frac{1}{2} \left( \frac{\mu}{r} \right)^{d-3} + O\left(1/r^{2(d-3)} \right) \, ,
\end{eqnarray}

We can interpret this geometry also as a wormhole connecting two asymptotic regions at $\rho= \pm 1$. The difference now is that we have some non-trivial red-shift between universes. In fact note that $g_{tt} \to 1$ when $\rho \to 1$, but $g_{tt} \to (1-2c_{\sigma})^2$ when $\rho \to -1$. This is similar to what happens in the Ellis wormhole \cite{Ellis:1973yv,Bronnikov:1973fh,Ellis:1979bh,Torii:2013xba,Blazquez-Salcedo:2018ipc}. Note that solution (\ref{wormhole_rho_2}) includes the solution (\ref{wormhole_rho}) in the limit $c_\sigma=0$.

Again the Ricci scalar of this solution is zero, but the Kretschmann scalar has a more complicated expression. In four dimensions we have
\begin{eqnarray}
	K = \frac{6(1-\rho^2)^6\left[1-2c_{\sigma}(1-\rho)+c_{\sigma}^2(2\rho^2-2\rho+1) \right]}{\mu^4[1-(1-\rho) c_{\sigma}]^2} \, . 
\end{eqnarray}
The Kretschmann scalar becomes singular at some radial point if $|1-1/c_{\sigma}|<1$.  This is actually the case for higher dimensions too. We can prevent the geometry from becoming sick if we choose $c_\sigma<1/2$.
Note that, if this expression holds, then $\rho\sigma>0$ everywhere. We have also assumed that $\kappa>0$, but different sign choices result in equivalent solutions, with some differences in the global signs of the parameters.

Let us calculate the mass of the wormhole using the Komar integral. {We need a time-like Killing vector normalized to one at infinity. For $\rho \rightarrow 1$ we can use the Killing $K^{+}_t=\partial_t$, because
	\begin{eqnarray}
	\lim\limits_{\rho \rightarrow +1} \mathbf g(\partial_t, \partial_t) = \lim\limits_{\rho \rightarrow +1} (\rho^2 \sigma^2) = 1
	\end{eqnarray}
	due to our normalization. The mass calculated using the dual form $\xi^{+}_t$ of the Killing $K^{+}_t$, using the expression 
	\begin{eqnarray}
	\mathbf d \xi^{+}_t &=& -2 c_\sigma \left[ 1- c_\sigma (1-\rho)  \right] \mathbf d t \wedge \mathbf d \rho \nonumber \\
	&=& -c_\sigma \mu^{d-3} (d-3) \left[ \frac{1-\rho^2}{\mu^{d-3}} \right]^\frac{d-2}{d-3} \boldsymbol{\omega}^t \wedge \boldsymbol{\omega}^r \, ,
	\end{eqnarray}
	is
	\begin{eqnarray}
	M_{+} &=&  - \frac{1}{16 \pi} \frac{d-2}{d-3} \int\limits_{\rho \rightarrow + 1} \ast \mathbf d \xi^{+}_t 
	= c_\sigma \frac{\mu^{d-3} (d-2) A_{d-2}}{16 \pi} \, . \;
	\end{eqnarray}

	For the other side we cannot use $\partial_t$, because
	\begin{eqnarray}
	\lim\limits_{\rho \rightarrow -1} \mathbf g(\partial_t, \partial_t) = (1 - 2 c_\sigma)^2 \, ,
	\end{eqnarray}
	which is generally not equal to one.  Instead, let us define the Killing vector
	\begin{eqnarray}
	\label{Killing_Bob}
	K^{-}_t = \frac{1}{|1 - 2 c_\sigma|} \partial_t \, .
	\end{eqnarray}
	Hence we have $\lim_{\rho \rightarrow -1} \mathbf g( K^{-}_t, K^{-}_t) = 1$. We also have to be careful with regard to the vielbein we use. Because $\mathbf e_r$ points towards spatial infinity on the $\rho > 0$ side, but it points towards the wormhole on the $\rho < 0$ side. So we have to reorient the vielbein for the mass calculation as well, changing from $\boldsymbol{\omega}^r$ to $-\boldsymbol{\omega}^r$. This introduces a minus sign in the star operator, and the expression of the Komar integral is
	\begin{eqnarray}
	M_{-} =  - \frac{1}{16 \pi} \frac{d-2}{d-3} \int\limits_{\rho \rightarrow - 1} (-\ast) \mathbf d \xi^{-}_t = -\frac{M_{+}}{|1 - 2 c_\sigma|} \, ,
	\end{eqnarray} 
	with $\xi^{-}_t$ the dual form of $K^{-}_t$. This relation indicates that each side measures values for the mass of the wormhole with contrary signs, but also with different absolute values. Note that the mass is finite in both sides, as long as $c_\sigma<1/2$ is satisfied.

	 If we insist in having a positive value for $M_{+}$, then $0 \le c_\sigma < 1/2$. In this case the value of $M_{-}$ is always negative.
Note also that for the singular solution with $c_\sigma=1/2$, the $M_-$ mass diverges but the $M_+$ mass reaches its maximum possible value. 

Note that we can find solutions with the opposite behaviour of the $\rho<0$ and $\rho>0$ sides if we choose a different sign of $\kappa$. In that case the $M_+$ mass could diverge while the $M_-$ mass would always remain finite.

	Before discussing in more detail the properties of the Dirac fields that support this wormhole geometry, let us explicitly calculate the position of the throat. For this we look at a slice of constant time and keep all angles constant except one. Then we embed this into a two-dimensional metric in a three-dimensional space using a function $z(r)$
\begin{eqnarray}
\mathrm d s_\text{embed}^2 &=& \mathrm d z^2 + \mathrm d r^2 + r^2 \mathrm d \phi^2  \\
&=& \left( \left[ \frac{\mathrm d z}{\mathrm d r} \right]^2 + 1 \right) \mathrm d r^2 + r^2 \mathrm d \phi^2 = \frac{1}{N} \mathrm d r^2 + r^2 \mathrm d \phi^2 \, , \nonumber
\end{eqnarray}
meaning
\begin{eqnarray}
\frac{\mathrm d z}{\mathrm d r} = \pm \sqrt{\frac{1}{N} - 1} \, .
\end{eqnarray}
The position of the throat can be calculated from $\mathrm d r/\mathrm d z |_{r_\text{throat}} = 0$, so from
\begin{eqnarray}
\left. \frac{N}{N-1} \right|_{r_\text{throat}} = 0
\end{eqnarray}
implying $r_\text{throat} = \mu$. 

\subsubsection{Properties of the unbounded spinors}

The spinor functions that support the wormhole geometry are
\begin{eqnarray}
f &=& \sqrt{\frac{d-3}{2 \mu \alpha_g}} \mathrm e^{c_\Delta / 2} \frac{-1}{\sqrt{1+ \frac{c_\sigma}{1-c_\sigma} \rho}} (1-\rho)^\frac{d-2}{d-3} \, , \nonumber \\
g &=& \sqrt{\frac{d-3}{2 \mu \alpha_g}} \mathrm e^{-c_\Delta / 2} \frac{1}{\sqrt{1+ \frac{c_\sigma}{1-c_\sigma} \rho}} (1+\rho)^\frac{d-2}{d-3} \, , 
\end{eqnarray}
To analyze the behavior of the matter content we look at the density $j^0_\text{net}$, which in this case looks like
\begin{eqnarray}
	j^0_\text{net} &=& \frac{d-3}{\alpha_g \mu} \frac{e^{c_\Delta}}{1+\frac{c_\sigma}{1-c_\sigma} \rho} \left[ (1-\rho)^{\frac{2 (d-2)}{d-3}} \mathrm  + (1+\rho)^{\frac{2 (d-2)}{d-3}} \mathrm e^{-2c_\Delta} \right] \, . \nonumber \\
\end{eqnarray}
with the relations (\ref{eqn:csigma}), (\ref{eqn:sigma2}), and $\kappa=(d-2)/2$.
%

The density $j^0_\text{net}$ is in general not zero in any of the asymptotic regions $\rho = \pm 1$:
\begin{eqnarray}
\label{j0_wh_inf}
	j_\text{net}^0(\rho=1) = \frac{d-3}{\alpha_g \mu} 
	2^{\frac{2 (d-2)}{d-3}}(1-c_\sigma)e^{-c_\Delta}  \nonumber \\
	j_\text{net}^0(\rho=-1) = \frac{d-3}{\alpha_g \mu} 
	2^{\frac{2 (d-2)}{d-3}}\frac{1-c_\sigma}{1-2c_\sigma}e^{c_\Delta} \, ,
\end{eqnarray}
meaning that the Dirac spinors are not bounded: the integral over all of the wormhole space-time of the Dirac density diverges, and hence the Dirac fields cannot be normalized.

From these expressions we can also see that the field density is regular and positive everywhere as long as $c_\sigma<1/2$, but like the Kretschmann scalar, it diverges on the left side when $c_\sigma=1/2$.

We show a few examples for the function $j_\text{net}^0$ in Figure \ref{fig:plotj0wormhole}. The minimum of the density in general does not coincide with the throat of the wormhole ($\rho=0$).

\begin{figure}
	\centering
	\includegraphics[width=0.75\linewidth,angle=-90]{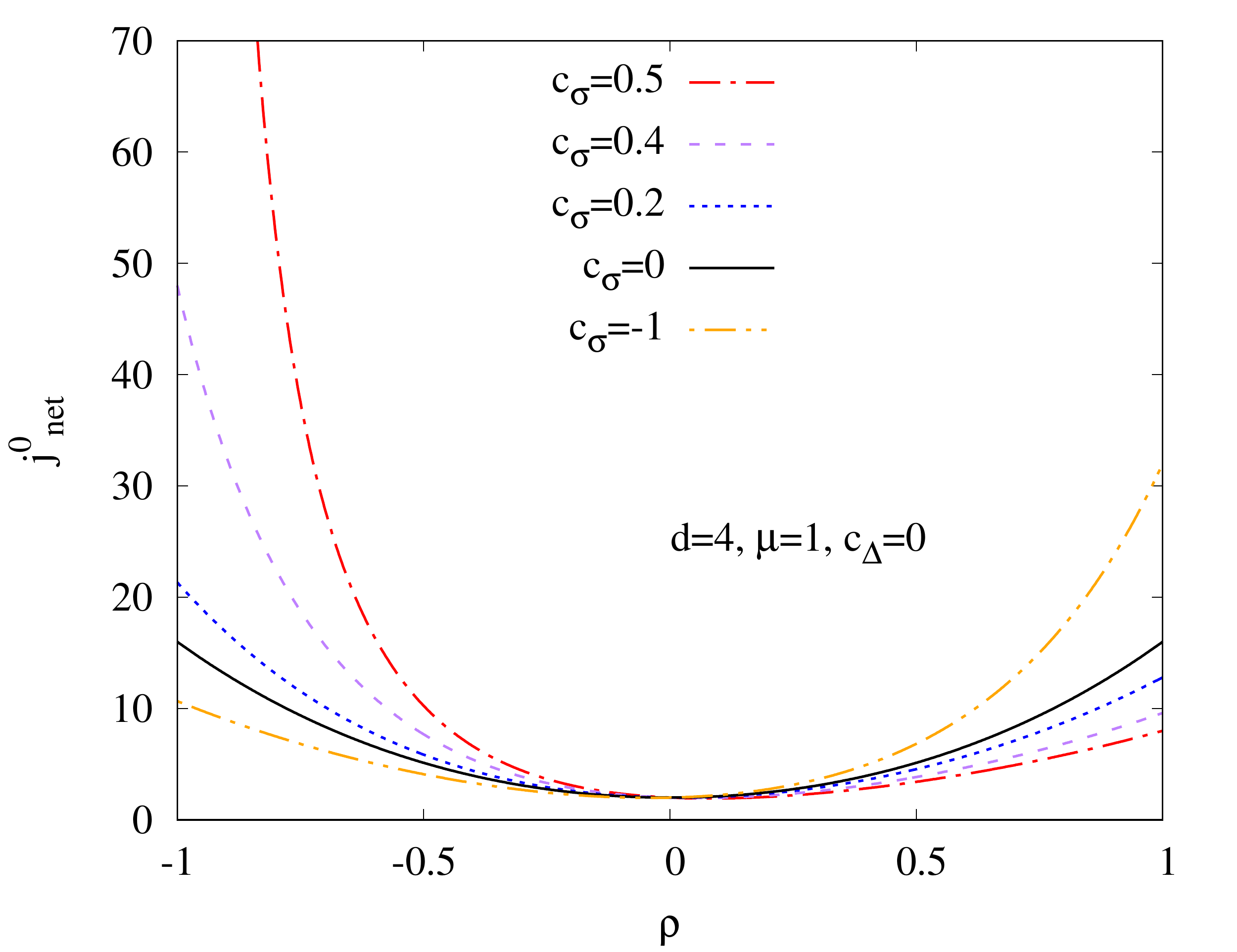}
	\caption{Dirac density $j^0_\text{net}$ as a function of the coordinate $\rho$ for $d=4$ wormholes. We choose $\mu=1$, $\alpha_g=1$, $c_\Delta=0$, and several values of the $c_\sigma$ parameter. 
	}
	\label{fig:plotj0wormhole}
\end{figure}

\begin{figure}
	\centering
	\includegraphics[width=0.75\linewidth,angle=-90]{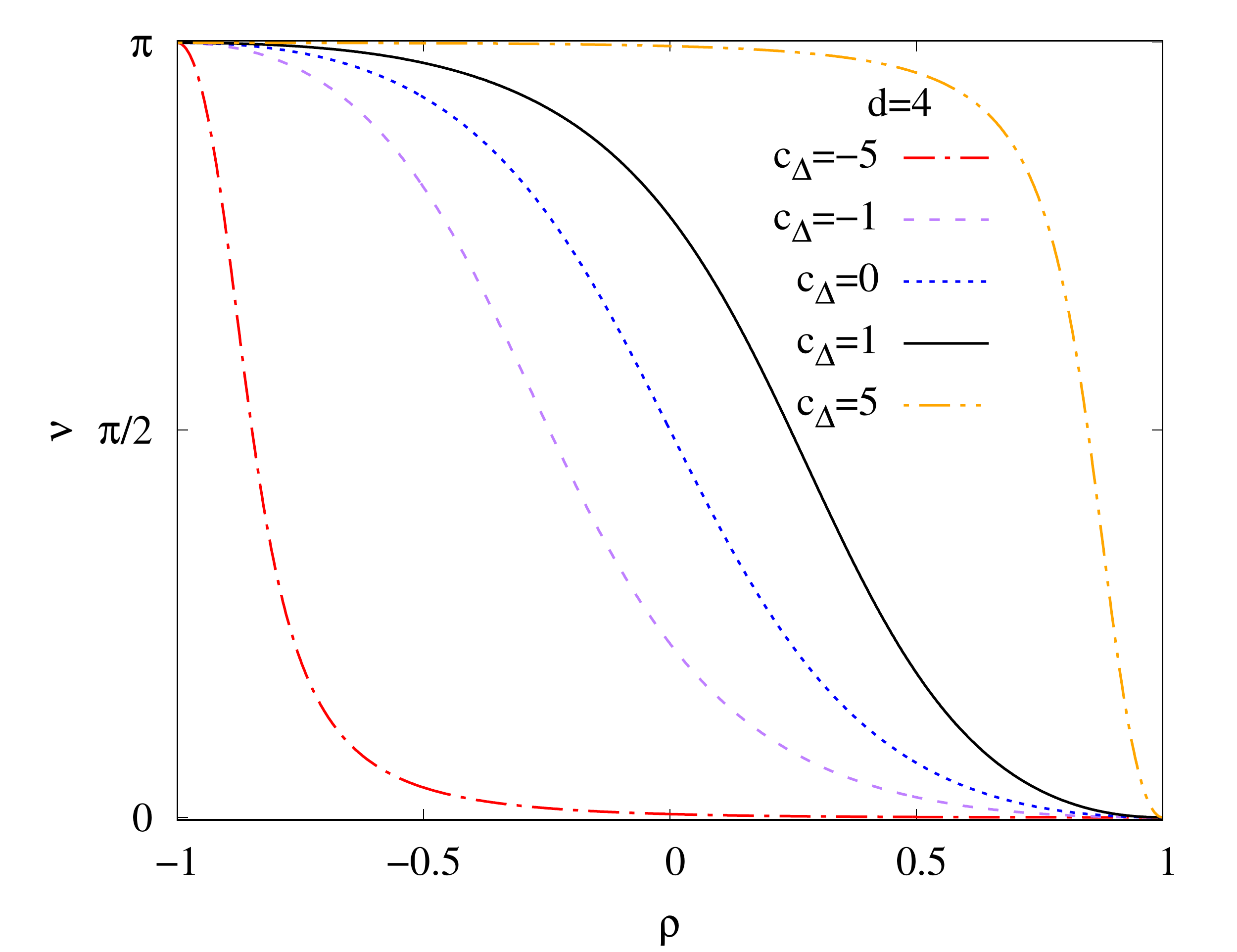}
	\caption{Phase $\nu$ as a function of the coordinate $\rho$ for $d=4$ wormholes. We choose several values of the $c_\Delta$ parameter. 
	}
	\label{fig:plotj0wormhole_phase}
\end{figure}

Keeping with $\kappa$ positive, notice that for $c_\sigma < 1/2$
\begin{eqnarray}
\label{ratio}
\frac{j^0_\text{net}(\rho = +1)}{j^0_\text{net}(\rho = -1)} = \frac{|M_+|}{|M_-|} \mathrm e^{-2 c_\Delta} \, ,
\end{eqnarray}
so $c_\Delta$ determines how much the ratio of the field amplitude in the asymptotic region differs from the ratio of the absolute values of the masses measured in these regions. 

To discuss the role of $c_\Delta$ further, let us look at the phase $\nu(\rho)$. From equation (\ref{eqn:set}) we have
\begin{eqnarray}
\mathrm e^{\mathrm i \nu} = \frac{(1+\rho)^{\frac{d-2}{d-3}} + \mathrm i (1 - \rho)^{\frac{d-2}{d-3}} \mathrm e^{c_\Delta}}{(1+\rho)^{\frac{d-2}{d-3}} - \mathrm i (1-\rho)^\frac{d-2}{d-3} \mathrm e^{c_\Delta}} \, .
\end{eqnarray}
In Figure \ref{fig:plotj0wormhole_phase} we show a plot of the phase function $\nu$ in $d=4$ for various values of $c_\Delta$. As we can see it changes from the boundary value $\nu(\rho = -1) = \pi$ to the boundary value $\nu(\rho=+1) = 0$ at a position determined by $c_\Delta$. For $c_\Delta = 0$ this phase jump happens at $\rho = 0$, for $c_\Delta < 0$ in the region $\rho < 0$ and for $c_\Delta > 0$ in the region $\rho > 0$. 

This change of phase suggests a relation between the spinors as defined by an asymptotically flat observer on the right side or on the left side. Following \cite{Cariglia:2018rhw} let us discuss the spinor field on both asymptotic flat regions. For this let us introduce the observers Alice and Bob. Alice will be the observer living in the asymptotically flat region $\rho \rightarrow 1$. Quantities like $M_{+}$ and $\kappa$ have been defined in the frame and with the vielbein of Alice.
 The observer Bob lives in $\rho \rightarrow -1$. Bob differs from Alice by his choice of time normalization (the temporal Killing vector Bob has to use is given by equation (\ref{Killing_Bob})) and by his choice of vielbein. Hence, if by $(A)$ we indicate Alice definitions and by $(B)$ Bob definitions, we can write
\begin{eqnarray}
t^{(B)} &=& (1 - 2 c_\sigma) t^{(A)} \, , \nonumber \\
\boldsymbol{\omega}^{r \, (B)} &=& - \boldsymbol{\omega}^{r \, (A)} \, .
\end{eqnarray}
We have already discussed how this changes the mass of the wormhole defined by Alice ($M^{(A)}=M_{+}$) and by Bob ($M^{(B)}=M_{-}$). But the change in the vielbein also affects the Dirac spinor.
 This change can be expressed as a unitary transformation
 changing $\gamma^r$ to $- \gamma^r$. To see the effect of this, let us write the radial part of the Dirac equation from the point of view of Alice
\begin{eqnarray}
\mathrm i \gamma^r \mathbf e^{(A)}_r \phi^{(A)} + \mathrm i \gamma^r  (\mathbf e^{(A)}_r F) \phi^{(A)} + \frac{\mathrm i}{r} \gamma^t \gamma^r \kappa^{(A)} \phi^{(A)} = 0 \, , 
\end{eqnarray}
with 
\begin{eqnarray}
F = \begin{cases} \ln \sqrt{ \sqrt{N(r)} \sigma(r) r^{d-2}} \; &, \; \text{for} \; \rho \ge 0 \\ \ln \sqrt{ \sqrt{N(r^\prime)} \sigma(r^\prime) r^{\prime \, d-2}} \; &, \; \text{for} \; \rho \le 0 \end{cases} \, ,
\end{eqnarray}
and the radial coordinate $r^\prime$ defined on the part $\rho < 0$ by the equation $-\rho = \sqrt{N(r^\prime)}$.

Changing to Bobs frame we have the Dirac equation
\begin{eqnarray}
\mathrm i \gamma^r \mathbf e^{(B)}_r \phi^{(B)} + \mathrm i \gamma^r  (\mathbf e^{(B)}_r F) \phi^{(B)} + \frac{\mathrm i}{r} \gamma^t \gamma^r (-\kappa^{(A)}) \phi^{(B)} = 0 \, , 
\end{eqnarray}
where we used $\mathbf e_r^{(B)} = - \mathbf e_r^{(A)}$. This shows us that
\begin{eqnarray}
\kappa^{(B)} = - \kappa^{(A)} \, ,
\end{eqnarray}
so the chirality of the field in Bob frame is the opposite to the chirality in Alice frame.

Thus we can write any of the spinors in the incoherent superposition of fields given by the sign choices $\boldsymbol{\epsilon}$ in the asymptotic regions as
\begin{eqnarray}
\Psi^{(A, B)}_{\kappa, \boldsymbol{\epsilon}}(|\rho|) = \Psi^{(A)}_{\kappa, \boldsymbol{\epsilon}} (r) \otimes \Psi^{(B)}_{-\kappa, \boldsymbol{\epsilon}}(r^\prime)
\end{eqnarray}
where we should keep in mind that the first part of that product is written in the frame of Alice and the second in the frame of Bob. So one could think of the spinors in the asymptotic regions as anticorrelated entangled pairs completely similiar to the discussion in \cite{Cariglia:2018rhw}. 
The change of chirality is a result of the different orientation of observers at each side of the wormhole, and it has nothing to do with the particular solution we have obtained, so this should be a general feature of Dirac fields in the geometry of wormholes. The only difference with other models is that this wormhole solution is a full back-reacting solution of the Einstein-Dirac equations, so the wormhole geometry is supported by the Dirac spinors.

	Let us comment now on the energy conditions of this wormhole. The stress-energy tensor for this particular solution is
	\begin{eqnarray}
		T_{tt} &=& 0 \, , \nonumber \\
		T_{rr} &=& - \frac{2 \kappa \mathrm e^{c_0}}{\sigma \rho r^{d-1}} \, , \nonumber \\
		T_{tj} &=& 0 = T_{r j} = T_{tr} \, , \nonumber \\
		T_{jk} &=& \frac{\mathrm e^{c_0}}{\sigma \rho r^{d-1}} \delta_{jk} \, .
	\end{eqnarray}
	So there are directions for which the energy conditions (null and weak) are violated. One could also easily see that from the fact that, due to a massless field $\tensor{T}{^a_a} = 0$ but also $T_{tt} = 0$, so there must be directions for which $T_{ab} \xi^a \xi^b < 0$ for time-like $\xi^a$. This is a generic feature of wormholes supported by exotic matter. In addition, in this case we have seen that the density of the Dirac fields does not decay at infinity, meaning the spinors are not bounded and it is also more difficult to interpret the Dirac fields of these solutions. 

	For instance, these wormholes we have obtained are described in practice by three parameters: $\mu$, $c_\sigma$ and $c_\Delta$ (apart from the coupling constant $\alpha_g$). Essentially, these three parameters are related respectively with the radius of the throat, the mass of the wormhole, and the amplitude of the Dirac fields. 
	In principle, the amplitude can be fixed by imposing an extra normalization condition on the field. 
	Since the Dirac fields are not bounded, it is not possible to fix the integral over the density to be equal to one, and hence we cannot make a quantum (probabilistic) interpretation of the field.
	One could fix this parameter following other criteria, for example, by fixing the value of the density (\ref{j0_wh_inf}) at one of the asymptotical regions, or by choosing a particular relation between the mass and density ratios (\ref{ratio}) .

Let us note that
the properties of Dirac fields in the background of wormhole geometries have been studied before in the literature \cite{Cariglia:2018rhw, Rojjanason:2018icy,Maldacena:2018gjk}. Solutions with pairs of Dirac fields 
in the background of a wormhole can be used as effective models describing a short nanotube bridging two different graphene layers. These models are known as graphene wormholes \cite{Gonzalez:2009je,Atanasov_2011,Pincak2013,Smotlacha:2014tza,Sepehri:2016svv,Sepehri:2017dky,Garcia:2019gro}. Such models are constructed in lower dimensions ($d=2+1$), in the presence of a gauge field and without back-reaction. In particular in \cite{Gonzalez:2009je} it was shown that, although there are test field configurations that cannot be normalized, under some conditions (vanishing angular momentum) bounded states of massles fermions localized around the wormhole can be constructed. 

%


For the analytical and back-reacting solution we have presented in this section, we have shown that the Dirac fields are not bounded, since the spinor functions are given by a combination of a part that goes to zero at infinity and a part that goes to a constant. A difference with the results from \cite{Gonzalez:2009je} is that our solution, spherical symmetry fixes the angular momentum to be $|\kappa|=(d-2)/2$.
There doesn't seem to be a simple generalization of our solution that results in bounded spinors, with a Dirac density decaying fast enough to zero at any of the asymptotical regions.

A natural question is to ask if these solutions can be generalized to include massive Dirac fields $m\neq0$ and/or frequency $\omega$, and if this could help obtain wormhole solutions with localized Dirac fields. Again, a simple analytical solution doesn't seem to be available, and this suggests that numerical methods may be necessary for the construction and analysis of these configurations.

However we can argue that a generalization of the previous $m=\omega=0$ wormhole solution are likely to result also in unbounded configurations.
Consider the case with $m\neq 0$, $\omega \neq 0$. If we assume an asymptotically flat space-time, the asymptotical behaviour of the massive (test) spinors is given by a combination of the form
\begin{eqnarray}
f = \frac{C_+}{r} e^{\sqrt{m^2-\omega^2}r}+ \frac{C_-}{r} e^{-\sqrt{m^2-\omega^2}r}\, , \nonumber \\
g = -\frac{C_+}{r}\frac{m+\omega}{m-\omega} e^{\sqrt{m^2-\omega^2}r}+ \frac{C_-}{r}\frac{m+\omega}{m-\omega} e^{-\sqrt{m^2-\omega^2}r}\, ,
\end{eqnarray}
provided $|\omega|<m$. Note that this asymptotical solution is given by a combination of a diverging term (with amplitude $C_+$) and a convergent term (with amplitude $C_-$). 
Hence it is reasonable to expect that the generalization of the previous wormhole solution to the case of massive spinors would be given asimptotically by a combination of such two contributions. This would result, in principle, in a configuration with unbounded spinors, with one of the functions decaying exponentially at infinity, while the other function explodes exponentially. Nonetheless, a full numerical construction of these solutions should allow for a regular and traversable throat, with a non-trivial matching of the spinor functions. Such a numerical analysis is beyond the scope of this paper, and it will be presented elsewhere. 

In this regard, in the case of black holes it has been proven that no fermionic bound states with $|\omega|<m$ exist \cite{Finster:1999ry,Finster:1998ak,Finster:1998ju,Kraniotis:2018zmh}. Nonetheless, a difference with respect the wormholes is that, in the case of black holes, the problem arises at the behaviour of the spinor functions at the horizon, where they diverge.
For the wormhole solution we have discussed in this section, as well as for other test field examples we have already mentioned \cite{Gonzalez:2018xrq}, the problem appears only asymptotically far from the throat.
  In the next section we are going to discuss another solution that exemplifies this point: a black hole for which the unbounded massless spinors diverges at the horizon.

\subsection{Schwarzschild black hole with a divergent Dirac flux}
\label{bh_section}

Let us now continue with the special case in which the spinors are massless with no frequency ($m=\omega=0$.). Consider the special solution for which either $f \equiv 0$ or $g \equiv 0$. In this case, from equations (\ref{wh_eqs}) we can see that the function $\sigma$ is just a constant, which we choose to be one. The metric is thus simply the $d$-dimensional Schwarzschild-Tangherlini metric. We will nevertheless have a non-vanishing spinor field in this background. The background is a vacuum black hole, because for the above spinor field configuration the stress-energy tensor vanishes. To have a well-behaved solution at infinity we choose $f = 0$ in the case of $\kappa < 0$ and $g=0$ in the case of $\kappa > 0$ (recall $\kappa$ is fixed by the Ansatz construction to $\kappa=\epsilon_\kappa \frac{d-2}{2}$). Define for this
\begin{eqnarray}
h = c_1 \exp \left[ \frac{- 2 |\kappa|}{d-3} \mathrm{artanh} \sqrt{N} \right] = \begin{cases} -f \; &\text{, for } \kappa > 0 \\ g \; &\text{, for } \kappa < 0 \end{cases} \, , \;\;\;\; 
\end{eqnarray}
with $c_1 \in \mathbb R_{\ge 0}$ being a constant.
This solution can actually be reached in the previous solution (\ref{sol_wh}), when taking in the expressions $c_\Delta=-\epsilon_\kappa c_0 +2\epsilon_\kappa\ln c_1$ and the limit $c_{\sigma}=1$ ($c_0 \to -\infty$).

The Dirac density of this solution is given by
\begin{eqnarray}
j^0_\text{net} = \frac{c_1^2}{r^{d-2}\sqrt{N}} \mathrm e^{-\frac{4 |\kappa|}{d-3} \mathrm{artanh} \left[ \sqrt{N} \right]} \, .
\end{eqnarray}
In Figure \ref{fig:plotj0blackhole} we show the Dirac density $j^0_\text{net}$ as a function of $r$ for several values of the dimension, where we can see that it decays to zero at infinity but diverges at the horizon. Let us analyze this behaviour in more detail.

The asymptotic part for determining the behaviour of the field at spatial infinity is
\begin{eqnarray}
\mathrm{artanh} \sqrt{y} \approx - \frac{1}{2} \ln(1-y) + \mathcal O( y^0) \, ,
\end{eqnarray}
for $y \le 1$ and $y \approx 1$. So for $r \rightarrow \infty$ we have
\begin{eqnarray}
\mathrm{artanh} \left[ \sqrt{1 - \left(\frac{\mu}{r}\right)^{d-3}} \right] \approx \frac{d-3}{2} \ln r + \mathcal O(r^0)
\end{eqnarray}
and thus
\begin{eqnarray}
j^0_\text{net} \approx \mathrm c_1^2 \frac{r^{-2 |\kappa|}}{r^{d-2}} = \frac{c_1^2}{r^{2(d-2)}} \rightarrow 0 \, .
\end{eqnarray}
Close to the horizon $j_\text{net}^0$ diverges like 
\begin{eqnarray}
j^0_\text{net} \approx 
\mathrm c_1^2 \frac{2(d-2) \mu^{2-d}}{\sqrt{d-3}\alpha_g}\frac{1}{\sqrt{r/\mu-1}}
+O(1)
 \, .
\end{eqnarray}
From this we can conclude that the integral over the density is not finite: the quantity
\begin{eqnarray}
\Sigma = \int \langle \mathbf j, \mathbf e_0 \rangle \mathrm d \Sigma^0 = A_{d-2} \int_{r_H}^\infty \frac{j^0_\text{net} r^{d-2}}{\sqrt{N}} \mathrm d r
\end{eqnarray}
explodes logarithmically at the horizon. 
Hence this solution is actually sick at the level of the matter field content, and is not physically reasonable. This is of course expected, since we have already mentioned that some general results forbid regular black holes with Dirac fields to exist \cite{Finster:1999ry,Finster:1998ak,Finster:1998ju,Kraniotis:2018zmh}.  Nonetheless, it is interesting to see that the combination of Dirac fields conspires in such a way so that the effective stress-energy tensor vanishes completely, and hence the geometry (the metric) is not affected at all by the matter configuration. Also it is interesting to see that the problem of the solution can be explicitly tracked to the behaviour of the Dirac fields at the horizon of the black hole.

\begin{figure}
	\centering
	\includegraphics[width=0.75\linewidth,angle=-90]{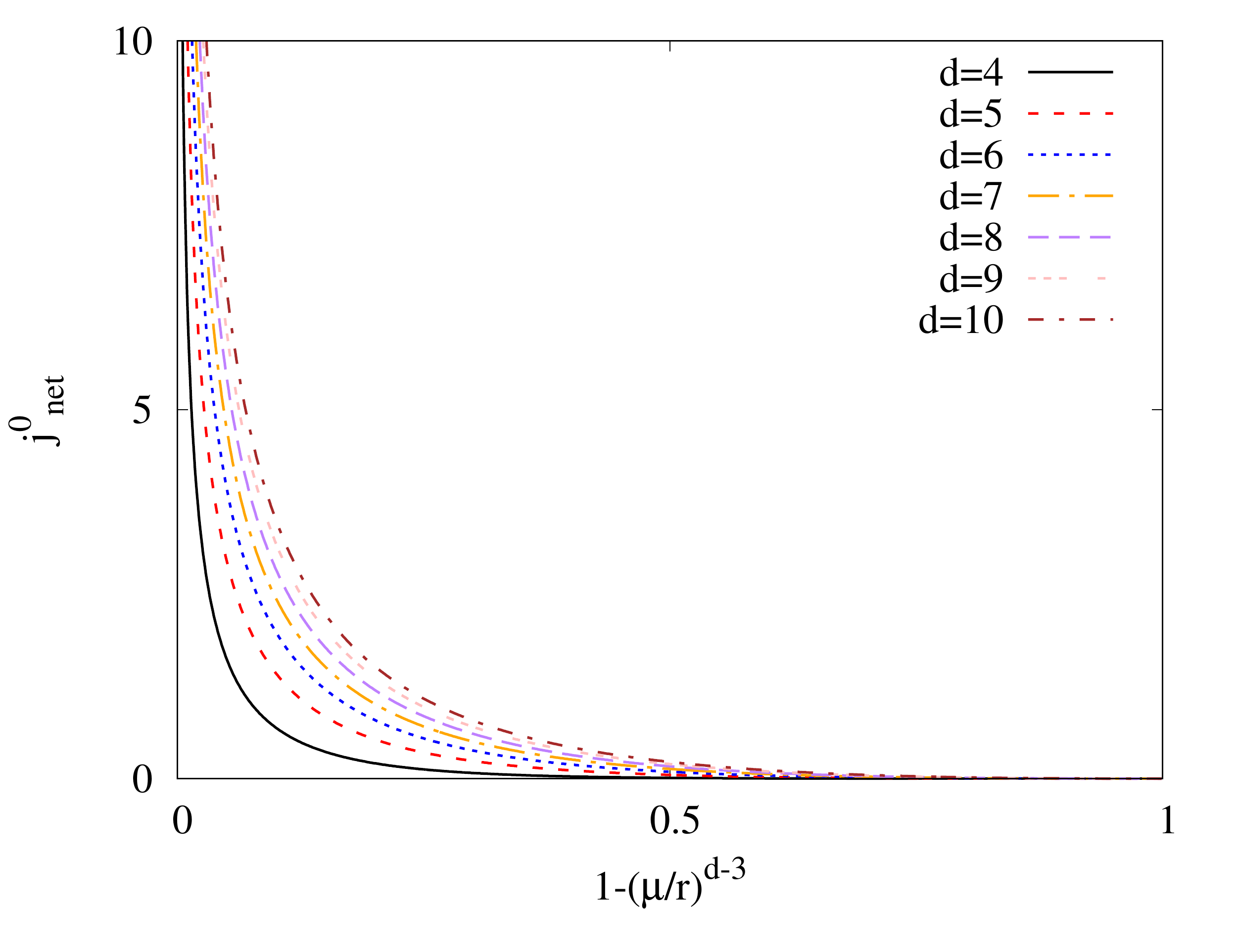}
	\caption{Dirac density $j^0_\text{net}$ as a function of the compactified radial coordinate $1-(\mu/r)^{d-3}$ for the black hole solution. Although the density decays fast enough at infinity, it diverges at the horizon. In the figure we choose $2\mathrm c_1^2 (d-2) \mu^{2-d}=\sqrt{d-3}\alpha_g$.}
	\label{fig:plotj0blackhole}
\end{figure}

\subsection{Light-like singularity}
\label{sing_section}

Another solution can be obtained when $m=0$ but $\omega \neq 0$. 
Specializing to solutions with $\frac{d\nu}{dr}=0$, we find that $\sigma$ has to be
\begin{eqnarray}
\sigma {=} - \frac{\omega}{\kappa \sin \nu} \frac{r}{\sqrt{N}} \, .
\end{eqnarray}
The equation (\ref{eqn:eq_of_motion_2}) for $\sigma$ implies that
 $N$ has to be a constant too, being
\begin{eqnarray}
N = \frac{d-3}{d-1} \, .
\end{eqnarray}
With this we can integrate the equation (\ref{eqn:eq_of_motion_2}) for $\lambda$, 
implying a simple expression for this function,
\begin{eqnarray}
\lambda = \frac{2 \kappa \cos \nu}{\sqrt{N}} \ln r/L \, ,
\end{eqnarray}
with $L \in \mathbb R$ a length scale defined from the integration constant. 

Lastly the equation (\ref{eqn:eq_of_motion_2}) 
implies two algebraic relations: 
\begin{eqnarray}
\label{algebraic_sing1}
0 &=& 2 \frac{d-3}{d-1} + \frac{\alpha_g (d-2) \sin^2 \nu}{4 \omega} L^{-(d-2)} \, ,  \\
-1 &=& -(d - 1) + \epsilon_\kappa (d-2) \sqrt{\frac{d-1}{d-3}} \cos \nu \, , \label{algebraic_sing2}
\end{eqnarray}
with $\kappa=\epsilon_\kappa(d-2)/2$.
Equation (\ref{algebraic_sing2}) fixes $\nu$
\begin{eqnarray}
\cos \nu = \epsilon_\kappa \sqrt{\frac{d-3}{d-1}} \, .
\end{eqnarray}
This can always be solved for $d \ge 3$.
With this
\begin{eqnarray}
\sin^2 \nu = 1 - \cos^2 \nu = \frac{2}{d-1}
\end{eqnarray}
and thus equation (\ref{algebraic_sing1}) implies the frequency has to be fine-tuned
\begin{eqnarray}
\omega = - \frac{\alpha_g \mathrm L^{-(d-2)}}{4} \frac{d-2}{d-3} \, .
\end{eqnarray}
Gathering all the relations, the solution is
\begin{eqnarray}
N &=& \frac{d-3}{d-1} \, , \nonumber \\
\sigma &=& \epsilon_\nu \epsilon_\kappa \frac{\alpha_g \mathrm L^{-(d-2)}}{2 \sqrt{2}} \frac{d-1}{d-3} \sqrt{\frac{1}{d-3}} \, r \, , \nonumber \\
\nu &=& \mathrm{arcsin} \sqrt{\frac{2}{d-1}} \, , \nonumber \\
\lambda &=& (d-2) \ln r/L \, ,
\end{eqnarray}
with $ \epsilon_\nu =\frac{\sqrt{d-1}}{2} \sin \nu = \pm 1 $. 
This means that
\begin{eqnarray}
j^0_\text{net} = \frac{2 \epsilon_\kappa \epsilon_\nu}{\alpha_g} \frac{d-3}{\sqrt{d-1}} \frac{1}{r} \, .
\end{eqnarray}
This fixes $\epsilon_\kappa \epsilon_\nu = +1$. The density of the field goes to zero at infinity, although it is divergent for $r \rightarrow 0$. 
We can simplify the expressions a bit by choosing the length $L^{-(d-2)}=\frac{2\sqrt{2}}{\alpha_g}\frac{d-3}{\sqrt{d-1}}$. The metric has a very simple form:
\begin{eqnarray}
\label{sing_metric0}
\mathrm d s^2 = r^2 \mathrm d t^{2} - \frac{d-1}{d-3} \mathrm d r^{2} - r^2 \mathrm d \Omega_{d-2}^2 \, ,  
\end{eqnarray}
The Kretschmann scalar for this solution in four dimensions $d=4$ is
\begin{eqnarray}
R^{abcd} R_{abcd} = \frac{8}{3 r^4} 
\end{eqnarray}
while the curvature scalar vanishes, $R=0$. The above singularity is light-like. To see this explicitly, we can calculate the Penrose diagram. The in- and out-going null geodesics obey the equation
\begin{eqnarray}
t = \pm \sqrt{\frac{d-1}{d-3}} \ln r + \text{const.} 
\end{eqnarray}
So making the following coordinate transformations
\begin{eqnarray}
\pi t^\prime &=&  \mathrm{arctan} \left(t + \sqrt{\tfrac{d-1}{d-3}} \ln r \right) +  \mathrm{arctan} \left(t - \sqrt{\tfrac{d-1}{d-3}} \ln r \right) \nonumber \\
\pi r^\prime &=&  \mathrm{arctan} \left( t - \sqrt{\tfrac{d-1}{d-3}} \ln r \right) -  \mathrm{arctan} \left(t + \sqrt{\tfrac{d-1}{d-3}} \ln r \right) \, , \nonumber \\
\end{eqnarray}
with $r^\prime, t^\prime  \in (-1, 1)$ brings the line element to
\begin{eqnarray}
\label{sing_metric} \nonumber \\
\mathrm d s^2 = \sqrt{\frac{d-3}{d-1}} r^2 (1+v^2)(1+w^2)(\mathrm d t^{\prime 2} - \mathrm d r^{\prime 2})- r^2 \mathrm d \Omega_{d-2}^2 \, , \nonumber \\
\end{eqnarray}
where $v=t+\sqrt{\frac{d-1}{d-3}}\ln r$ and $w=t-\sqrt{\frac{d-1}{d-3}}\ln r$.
The singularity is at $r \rightarrow 0$, meaning it is at $t^\prime + r^\prime = -1$ and $t^\prime - r^\prime = 1$.
This set defines a light-like surface. 
In Figure \ref{fig:plotsingularity} we show the Penrose diagram of metric (\ref{sing_metric}) in the $t^\prime$, $r^\prime$ coordinates and showing a few curves with constant $t$ and $r$.
All the radial geodesics of a massive particle begin and end at the singularity $r=0$.

\begin{figure}
	\centering
	\includegraphics[width=0.75\linewidth,angle=-90]{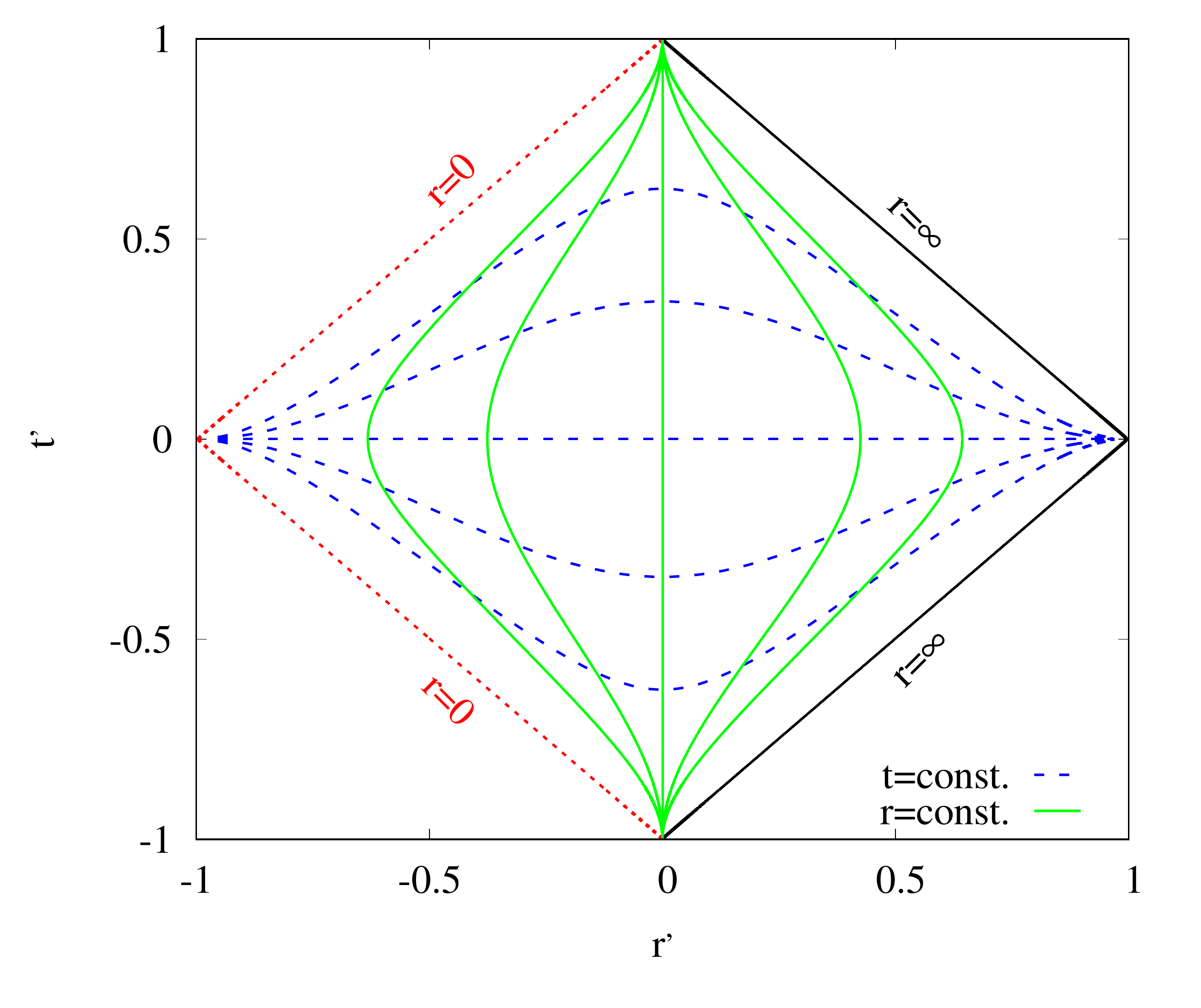}
	\caption{Penrose diagram for the metric (\ref{sing_metric}). Blue dotted lines show surfaces of constant time $t$. Green dashed lines show surfaces of constant radius $r$. 
	}
	\label{fig:plotsingularity}
\end{figure}

\section{Conclusions}
\label{section_conclusions}

In this paper we have studied the properties of configurations with a collection of Dirac fields, chosen in such a way that the total stress-energy tensor of the matter content is compatible with the spherical symmetry of the metric.

In sections \ref{sec:vielbein} and \ref{section_full_ansatz} we have given a detailed explanation on how the collection of $2^{\lfloor \frac{d-2}{2} \rfloor}$ Dirac fields can be chosen in order to achieve this symmetry. In order to do this, one proposes the standard separable Ansatz for each indivitual spinor. The radial and temporal dependence of the spinor is assumed to be equal for all fields. We make use of the known solutions for the angular part of the spinor, in particular when the $(d-2)$-sphere is factorized as a tower of lower dimensional spheres. Then it is possible to show that, for certain values of the angular momentum of the field, $|\kappa|=(d-2)/2$, the angular dependence of each field combines with the rest in a way so that the total stress-energy tensor is compatible with the spherical symmetry of the background. In order to have static metrics, we have to impose the vanishing of the radial current.

Making use of this Ansatz, we simplified the action and field equations in section \ref{section_action}. With this simplified system we constructed several simple analytical solutions.
In section \ref{section_wormhole} we obtained a family of wormholes supported by the Dirac fields. These wormholes connect two asymptotically flat regions, they can have positive, zero or negative mass, and their geometry and matter content are regular everywhere. However the density, although regular everywhere, does not decay to zero at any of the asymptotically flat regions.
We analyzed the relation between several quantities at each asymptotically flat region, in particular, the mass, which changes sign and value at each side, and the chirality of the Dirac fields, which also changes sign. 

In section \ref{bh_section} we found that the Schwarzschild metric can in fact satisfy the Einstein-Dirac field equations with a non-trivial solution for the fermionic fields. The catch is that the matter content becomes sick at the horizon of the black hole (the density diverges), and the fields cannot be normalized in the standard way.

In section \ref{sing_section} we have also obtained a light-like naked singularity, where matter fields and geometry become simultaneously sick.

For a study of more general solutions, it is likely that numerical methods are required in order to construct the solutions. For instance, the Ansatz we have presented here in detail was used to obtain numerically regular self-gravitating solitons 
\cite{Blazquez-Salcedo:2019qrz}.

A possible continuation of this work is to investigate numerically if it is possible to obtain similar wormhole solutions with finite values of the Dirac mass and/or frequency, or with additional matter fields (like a gauge field, similar to what is used in the graphene wormholes). It would be interesting to explore if in these cases regular solutions still exist, and more importantly, if bounded states can be constructed, with the Dirac density asymptotically zero.  On the other hand, these non-trivial asymptotics of the Dirac fields in the case of the wormhole, suggest that maybe allowing the metric to have other asymptotical behaviour (i.e., allowing for a negative cosmological constant to have an asymptotically AdS space-time) could help regularize the integral of the Dirac density.  

\begin{acknowledgements}
The authors would like to thank Prof. Dr. Jutta Kunz and Dr. Eugen Radu for helpful discussions and comments on this manuscript. 
JLBS and CK would like to acknowledge support by the
DFG Research Training Group 1620 {\sl Models of Gravity}.
JLBS would like to acknowledge support from the DFG project BL 1553, and acknowledge networking support by the
COST Action CA16104 {\sl GWverse}. 
\end{acknowledgements}

\bibliography{paperV0}

\end{document}